\documentclass[superscriptaddress,onecolumn]{revtex4}
\usepackage{graphicx,epsfig}
\usepackage{amsmath}
\usepackage {amssymb}
\usepackage{float}
\usepackage[usenames]{color}
\usepackage{soul}

\newcommand{\be}{\begin{eqnarray}}
\newcommand{\ee}{\end{eqnarray}}
\newcommand{\bea}{\begin{eqnarray}}

\newcommand{\eea}{\end{eqnarray}}

\def\C{{\cal{C}}}

\makeindex

\begin{document}

\title{Massive scalar field perturbations in Weyl black holes}

\author{Ram\'{o}n B\'{e}car}
\email{rbecar@uct.cl}
\affiliation{Departamento de Ciencias Matem\'{a}ticas y F\'{\i}sicas, Universidad Catolica de Temuco}

\author{P. A. Gonz\'{a}lez}
\email{pablo.gonzalez@udp.cl} \affiliation{Facultad de
Ingenier\'{i}a y Ciencias, Universidad Diego Portales, Avenida Ej\'{e}rcito
Libertador 441, Casilla 298-V, Santiago, Chile.}

\author{Felipe Moncada}
\email{fmoncada@uct.cl}
\affiliation{Departamento de Ciencias Matem\'{a}ticas y F\'{\i}sicas, Universidad Catolica de Temuco}

\author{Yerko V\'asquez}
\email{yvasquez@userena.cl}
\affiliation{Departamento de F\'isica, Facultad de Ciencias, Universidad de La Serena,\\
Avenida Cisternas 1200, La Serena, Chile.}

\date{\today}

\begin{abstract}

In this work we consider the propagation of  massive scalar fields in the background of Weyl black holes and we mainly study the effect of the scalar field mass in the spectrum of the quasinormal frequencies (QNFs) via the Wentzel-Kramers-Brillouin (WKB) method, and the pseudospectral Chebyshev method. The spectrum of QNFs is described by two families of modes, one of them is the photon sphere and the other one is the de Sitter. Essentially we show, via the WKB method, that the photon sphere modes exhibit an anomalous behaviour of the decay rate of the QNFs, that is, the longest-lived modes are the ones with higher angular number, as well as, there is a critical value of the scalar field mass, beyond which the anomalous behaviour is inverted. Also, we analyze the effect of the scalar field mass on each family of modes, and on their dominance, 
and we give an estimate value of the scalar field mass where the interchange in the dominance family occurs.

\end{abstract}

\maketitle

\tableofcontents

\clearpage

\section{Introduction}

In 1918 Hermann Weyl \cite{weyl:1918,weyl:1918b} attempted to unify the theory of General Relativity (GR) with electromagnetism. In this theory of gravity, the metric would transform under a conformal transformation $g_{\mu\nu}\rightarrow \Omega^{2}(x)g_{\mu \nu}$ whenever the electromagnetic field undergoes a gauge transformation $A_{\mu}\rightarrow A_{\mu}-\partial_{\mu}\log(\Omega(x))$, where $\Omega(x)$ is the local spacetime stretching, as a consequence the covariant derivative in Weyl's theory no longer preserve the metric, for that reason Weyl's theory of gravity never became a serious competitor for GR. However, Bach \cite{Bach:1921} derived a different theory of conformal gravity in 1921, whose action is constructed from contractions and squares of the Weyl tensor $\C_{\mu\nu\rho\sigma}$ in four dimensions which is conformally invariant and it is usually called Weyl or Weyl-squared gravity. It is important to point out that in four dimensions the Weyl-squared action is the unique conformally invariant action constructed solely from the Weyl tensor. On the other hand, the action of the theory gives rise to fourth-order equations of motion for the gravitational field which make difficult to reconcile it with Newtonian gravity.
One of the immediate consequences of postulating a gravitational theory with conformal invariance is that the artificially implanted cosmological constant, $\Lambda$, present in the Einstein-Hilbert action must be withdrawn, since not to do so would introduce a length scale that breaks the conformal symmetry of the theory. However, the same term naturally will emerge out of the metric, which provides further circumstantial evidence for the effectiveness of the principle under consideration.\\

Despite the absolute success of the GR theory,  it fail to describe observations on scales much higher than the solar system without placing a large amount of dark matter; however, the absence of any direct experimental evidence for dark matter \cite{Feng:2010gw}, has led to the consideration of various modified theories of gravity among which is conformal Weyl gravity, which may not require a dark matter component to explain the astrophysical data. 
The static and spherically symmetric vacuum solution describing a black hole was obtained by Mannheim and Kazanas \cite{Mannheim1989}, where a particular parameter of the solution can explain the flat rotation of galaxies without introducing dark matter. Also, the theory is intended to cover the dark energy related phenomena \cite{Mannheim:2005bfa,Nesbet:2012sk}. Moreover,
it was found three new exact solutions of this four-order theory, namely the Reissner-Nordstr\"om, Kerr and Kerr-Newmann solutions \cite{kazanas1991}. Other solutions of the conformal Weyl gravity can be found in \cite{Klemm:1998kf,Dzhunushaliev:1999fy,Said:2012,Lu:2012xu,Yang:2022icz, Herrera:2017ztd}.
 \\

Moreover, in the context of the detection of gravitational waves \cite{LIGOScientific:2016aoc}, the detected signal is consistent with GR 
\cite{LIGOScientific:2016lio}.  However, there are possibilities for alternative theories of gravity due to the large uncertainties in mass and angular momenta of the ringing black hole \cite{Konoplya:2016pmh}. So,  the study of the quasinormal modes (QNMs) and quasinormal frequencies (QNFs)
\cite{Regge:1957td,Zerilli:1970wzz,Kokkotas:1999bd, Nollert:1999ji,Konoplya:2011qq,Berti:2009kk} nowadays plays an important role.
The QNMs and QNFs give
information about the stability of matter fields that
evolve perturbatively in the exterior region of a black hole without backreacting on the
metric. Also, the QNMs are characterized by a spectrum that is independent of the initial conditions of the perturbation and depends on the black hole parameters, on the probe field parameters, and on the fundamental constants of the system. The QNFs are an infinite discrete spectrum of complex frequencies,
in which the real part 
determines the oscillation timescale of the modes, while the complex part
determines their exponential decaying timescale,
for a review on QNMs see \cite{Kokkotas:1999bd, Berti:2009kk}.\\

The tensor QNM spectrum for the Schwarzschild and Kerr black hole backgrounds show that the longest-lived modes are always the ones with lower angular number. This can be understood from the fact that  the more energetic modes with high angular number 
would have faster decaying rates. However, for the propagation of massive scalar field a different behaviour  was found
\cite{Konoplya:2004wg, Konoplya:2006br,Dolan:2007mj, Tattersall:2018nve},  at least for the fundamental mode. 
If the mass of the scalar field is light, then the longest-lived QNMs are those with a high angular number, 
whereas if the mass of the scalar field is large  the longest-lived modes are those with a low angular number. 
This behaviour, knowing as anomalous decay rate, is expected since if the probe scalar field is massive its fluctuations can maintain the QNMs to live longer even if the angular number 
is large. This behaviour of the QNMs introduces an anomaly of the decaying modes which depends on whether the mass of the scalar field exceeds a critical value or not. So, by introducing another scale in the theory through the presence of a cosmological constant an anomalous behaviour of QNMs was found in \cite{Aragon:2020tvq} as the result of the  interplay of the mass of the scalar field and the value of the cosmological constant. Anomalous decay rate of the QNMs  were also found if the background metric is the Reissner-Nordstr\"om and the probe scalar field is massive \cite{Fontana:2020syy} or massive and charged \cite{Gonzalez:2022upu} depending on critical values of the charge of the black hole, the charge of the scalar field and its mass. The presence of the anomalous behavior for a generalized Bronnikov-Ellis womhole and the Morris-Thorne wormhole was studied in Ref. \cite{Gonzalez:2022ote}. The anomalous decay rate of the QNMs has been  studied in various setups, see \cite{Aragon:2020teq,Destounis:2020pjk, Aragon:2020xtm, Aragon:2021ogo, Becar:2022wcj, Konoplya:2020fwg}.\\

The aim of this work is to study the propagation of massive scalar fields in Weyl black hole backgrounds, in order to study, the effect of the scalar field mass in such propagation. Some issues that we will address are the anomalous decay rate of QNMs, as well as, if there is a critical scalar field mass. Also, we will study the dominance between the family of modes, in order to analyze, if the dominant family suffers of such anomalous behaviour. It is worth mentioning that massless scalar field perturbations, dynamical  evolution  and Hawking radiation were recently studied for this spacetime, and it was shown that the propagation of massless scalar fields is stable. Also, the dominance between the two family of modes depending of the parameter $\lambda$ was established, being the photon sphere (PS) modes dominant for small $\lambda$. However, as $\lambda$ increases, the imaginary part of the PS mode decreases, whereas the de Sitter (dS) mode rises \cite{Fu:2022cul}. The parameter $\lambda$ has been considered in dark energy scenarios, and plays a role as the inverse proportion of the cosmological constant. Besides, the behaviour of the QNMs was used to study the thermal stability of black holes in conformal Weyl gravity, comparing this results with Schwarzschild black holes \cite{Momennia:2018hsm}. In the particular case of nearly extreme black hole in Weyl gravity, it was shown in \cite{Momennia:2019cfd} the correspondence between the parameters of the circular null geodesic and the QNFs in the eikonal limit, 
%$\ell \rightarrow \infty$, 
and the QNMs of the gravitational and electromagnetic perturbations on a black hole in (exact) Weyl gravity was calculated and studied in \cite{Momennia:2019edt}.\\

The manuscript is organized as follows: In Sec. \ref{background} we give a brief review about the Weyl black holes. Then, in Sec \ref{MSP}, we study massive scalar perturbations in the background of Weyl black holes. In Sec. \ref{QNM} we consider the PS modes and we find the critical scalar field mass. Also, we show the anomalous behaviour of the decay rate by using the WKB method. Then, we analyze the dS modes via the pseudospectral Chebyshev method, and we study the dominance family modes. Finally, we conclude in Sec. \ref{conclusion}.

\section{Four-dimensional Weyl black holes}
\label{background}

The action of four-dimensional Weyl gravity is given by \cite{Mannheim1989} 
 \begin{eqnarray} \label{action}
 S=-\alpha\int d^{4}x\sqrt{-g}C_{\mu\nu\rho\sigma}C^{\mu\nu\rho\sigma}+ I_{M}~,
\end{eqnarray}
where $\alpha$ is a dimensionless gravitational coupling constant which is usually chosen to be positive in order to satisfy the Newtonian lower limit, $I_{M}$ is the matter part of the action and $C_{\mu\nu\rho\sigma}$ is the Weyl tensor given by
\begin{equation}
    C_{\mu\nu\rho\sigma}=R_{\mu\nu\rho\sigma}+\frac{R}{6}\left(g_{\mu\rho}g_{\nu\sigma}-g_{\mu\sigma}g_{\nu\rho}\right)-\frac{1}{2}\left(g_{\mu\rho}R_{\nu\sigma}-g_{\mu\sigma}R_{\nu\rho}-g_{\nu\rho}R_{\mu\sigma}+g_{\nu\sigma}R_{\mu\rho}\right)\,,
\end{equation}
which satisfies the conformal invariance condition
\begin{equation}
  C_{\mu\nu\rho\sigma}\rightarrow \tilde{C}_{\mu\nu\rho\sigma}=\Omega^{2}(x)C_{\mu\nu\rho\sigma} \,. 
\end{equation}
By using the definition of the Weyl tensor and making use of the Gauss-Bonnet theorem it is possible to express the action (\ref{action}) in the following form
\begin{equation}
   S=-\alpha\int d^{4}x\sqrt{-g}\left(R^{\mu\nu}R_{\mu\nu}-\frac{1}{3}R^{2}\right)+ I_{M} \,.
\end{equation}
This theory of gravity is governed by field equations, %combined of second order differentiations of the Ricci scalar $R$. These equations, 
that can be derived by the functional variation of the action with respect to the metric $g_{\mu\nu}$ and take the following form:
\begin{eqnarray}
\label{bach}
\notag W_{\mu\nu} &=& 2C^\rho\,_{\mu\nu}\,{}^\sigma{}_{;\alpha\sigma}+C^\rho\,_{\mu\nu}{}^{\sigma}R_{\rho\sigma}
=\nabla^{\rho}\nabla_{\mu}R_{\nu\rho}+\nabla^{\rho}\nabla_{\nu}R_{\mu\rho}-\square R_{\mu\nu}-g_{\mu\nu}\nabla_{\rho}\nabla_{\sigma}R^{\rho\sigma}-2R_{\rho\nu}R^{\rho}_{\mu}+\frac{1}{2}g_{\mu\nu}R_{\rho\sigma}R^{\rho\sigma}\\
&& -\frac{1}{3}\left(2\nabla_{\mu}\nabla_{\nu}R-2g_{\mu\nu}\square R-2R R_{\mu\nu}+\frac{1}{2}g_{\mu\nu}R^{2}\right)=\frac{1}{4\pi}T_{\mu\nu}\,,
\end{eqnarray}
where $W_{\rho\sigma}$ is the Bach tensor.
It is important to note from (\ref{bach}) that in vacuum $T_{\mu\nu}=0$ $(W_{\mu\nu}=0)$ every solution $R_{\mu\nu}=0$ in Einstein-Hilbert action also leads to a solution in Weyl gravity; however,  not every vacuum solution from Weyl gravity implies a solution for GR.
The first static and spherically symmetric vacuum solution
describing a black hole in this theory was obtained by
Mannheim and Kazanas \cite{Mannheim1989}. The lapse function, used in the line element, is given by:
\begin{equation}\label{lapseweyl}
    f(r)=1-3\beta\gamma-\frac{2\beta-3\gamma\beta^{2}}{r}+\gamma r-k r^{2}\,,
\end{equation}
where the parameters $\beta$, $\gamma$ and $k$ are integration constants. In this solution,
the parameter $\gamma$ measures the departure of Weyl theory from GR, and so for small enough, both theories have similar predictions. On the other hand, it was argued that Weyl gravity can explain the flat rotation of galaxies without introducing dark matter, for which $\gamma$  
%is of
must be of the order of the inverse of the Hubble radius ($\gamma\approx \frac{1}{R_{H}}$) \cite{Mannheim1989}. Later, the solution (\ref{lapseweyl}) was generalized for rotating and charged solutions \cite{kazanas1991}. It is important to point out that this solution reduces to Schwarzschild black holes when $k=\gamma=0$ and to Schwarzschild-de Sitter black holes if $\gamma=0$.\\

In \cite{kazanas1991} the authors extended their first work \cite{Mannheim1989} and  presented the exact solution to the Reissner-Nordstr\"om problem associated with static, spherically  symmetric point electric and/or magnetic charge coupled to Weyl gravity. The metric function for the electric case looks like the following
\begin{equation}
  f(r)= 1-3\beta\gamma-\frac{2\beta-3\gamma\beta^{2}}{r}+\gamma r-k r^{2}-\frac{Q^{2}}{8 r\alpha\gamma} 
\end{equation}
where $Q$ is the electric charge. As pointed out by the authors, the first principal difference with the Reissner Nordstr\"om solution in standard Einstein theory is that the effect of the electromagnetic energy of a point electric charge is to produce a $1/r$ term in the exterior geometry of the black holes rather than the $1/r^2$ term present in GR. The second difference is that the geometry is not asymptotically conformally flat.  \\

In this work we follow Refs. \cite{Payandeh:2012mj,Fathi:2020sey} in which the authors applied the background field method in the weak field limit, and it was possible to derive other Reissner-Nordstr\"om solutions. The authors in \cite{Payandeh:2012mj} found a general metric solution given by the line element: 
\begin{eqnarray}
 ds^{2}=-f(r)dt^{2}+f^{-1}(r)dr^{2}+r^2 d \Omega^2\,, \label{metricBH}
 \end{eqnarray}
where $d\Omega^{2}$ is the line elements of the 2-sphere and the lapse function:
\begin{equation}
    f(r)=1+\dfrac{1}{3}(c_{2}r + c_{1}r^2)  \,, \label{f1}
\end{equation}
the coefficients $c_1$ and $c_2$ were found using the method named above. The last two terms of this $f(r)$ can be seen as a perturbation to the Minkowski spacetime ($h_{\mu\nu}=g_{\mu\nu}-\eta_{\mu\nu}$), which was studied using the Poisson equation $\nabla ^2 h_{\mu \nu}= 8\pi \mathcal{T}_{\mu \nu}$. Now using the weak field limit (zero-zero component):
\begin{equation}
    \nabla ^2 h_{00}= 8\pi( T_{00}+E_{00})= 8\pi \left(\dfrac{m_0}{\frac{4}{3}\pi r_0^3} + \dfrac{1}{8\pi} \dfrac{q_0^2}{r^4}\right). \label{h00}
\end{equation}
Here, $T_{00}$ is the scalar part of the energy-momentum tensor of a source of mass $m_0$, radius $r_0$ and $E_{00}$ is the another part of the energy-momentum tensor associated to the charge amount $q_0$ of the massive source. Using (\ref{f1}) in  (\ref{h00}) it was obtained:
\begin{equation}
    c_2 = -\dfrac{9 r m_0}{r_0^3} - \dfrac{3q_0^2}{2r^2} -3c_1 r\,, \label{c2}
\end{equation}
and substituting this in (\ref{f1}) it was found
\begin{equation}
 f(r)=1-\frac{r^2}{\lambda^{2}}-\frac{Q^2}{4r^2}\,, \label{f(r)}
 \end{equation} 
where
\begin{eqnarray}
    \dfrac{1}{\lambda^2} = \dfrac{3m_0}{r_0^3} + \dfrac{2c_1}{3}~,\,\, \text{and} \,\, Q = \sqrt{2}q_0 \,.
\end{eqnarray}

  It should be noted its attractive inverse square potential due to the charged body, instead of the repulsive in Reissner-Nordstr\"om-de Sitter black hole. 
For $\lambda>Q$, the roots of the lapse function are 
\begin{eqnarray}
\label{rh}
    r_{h}=\lambda\sin{(\frac{1}{2}\arcsin{(\frac{Q}{\lambda})})}~,\,\, \text{and} \,\, r_{c}=\lambda\cos{(\frac{1}{2}\arcsin{(\frac{Q}{\lambda})})}~.\,\,
\end{eqnarray}

\begin{figure}[h]
\begin{center}
\includegraphics[width=0.5\textwidth]{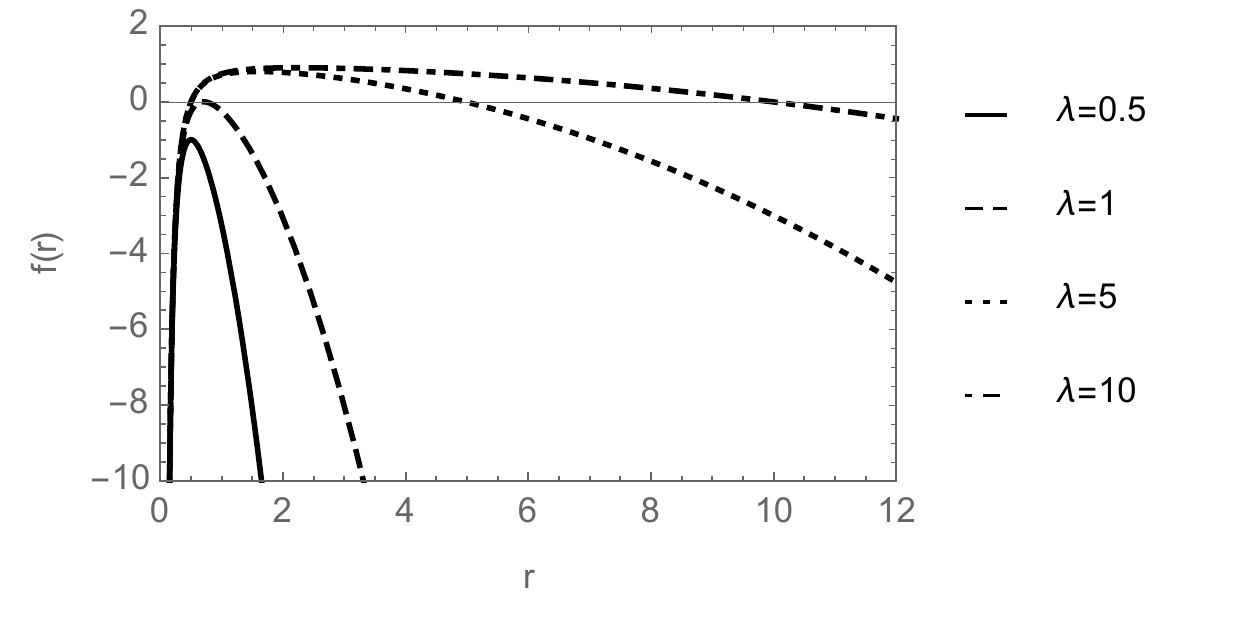}
\end{center}
\caption{The behaviour of $f(r)$ with $Q=1$, and different values of $\lambda$.}
\label{Lapse}
\end{figure}

The extremal black hole is obtained when $\lambda=Q$ and both horizons coalesce to $r_{ext}=\frac{\lambda}{\sqrt{2}}$. On the other hand, for $\lambda<Q$ a naked singularity is encountered. In Fig. \ref{Lapse} we plot the lapse function, where for a fixed value of the black hole charge $Q=1$, it is possible to observe the transition among a naked singularity $\lambda=0.5$, the extremal black hole $\lambda=1$, and a black hole with two horizons $\lambda > 1$. 

%\newpage

\section{Massive scalar field perturbations}
\label{MSP}

In order to obtain the QNMs of scalar field perturbations in the background of  the metric (\ref{f(r)})
we consider the Klein-Gordon equation
\begin{equation}
\label{KGE}
\Box \psi = \frac{1}{\sqrt{-g}}\partial_{\mu}\left(\sqrt{-g} g^{\mu\nu}\partial_{\nu}\right)\psi=m^2\psi~,
\end{equation}
with suitable boundary conditions, that is,  only ingoing waves on the horizon, and on the cosmological horizon.
%for a black hole geometry. 
In the expression above $m$ is the mass of the scalar field $\psi $.  It is worth mentioning that the Weyl tensor is traceless which implies that the stress-energy tensor of the matter fields must be traceless too. However, for a probe field it is not actually necessary to respect the same symmetries of the Weyl tensor.\\

\begin{figure}[h]
\begin{center}
\includegraphics[width=0.45\textwidth]{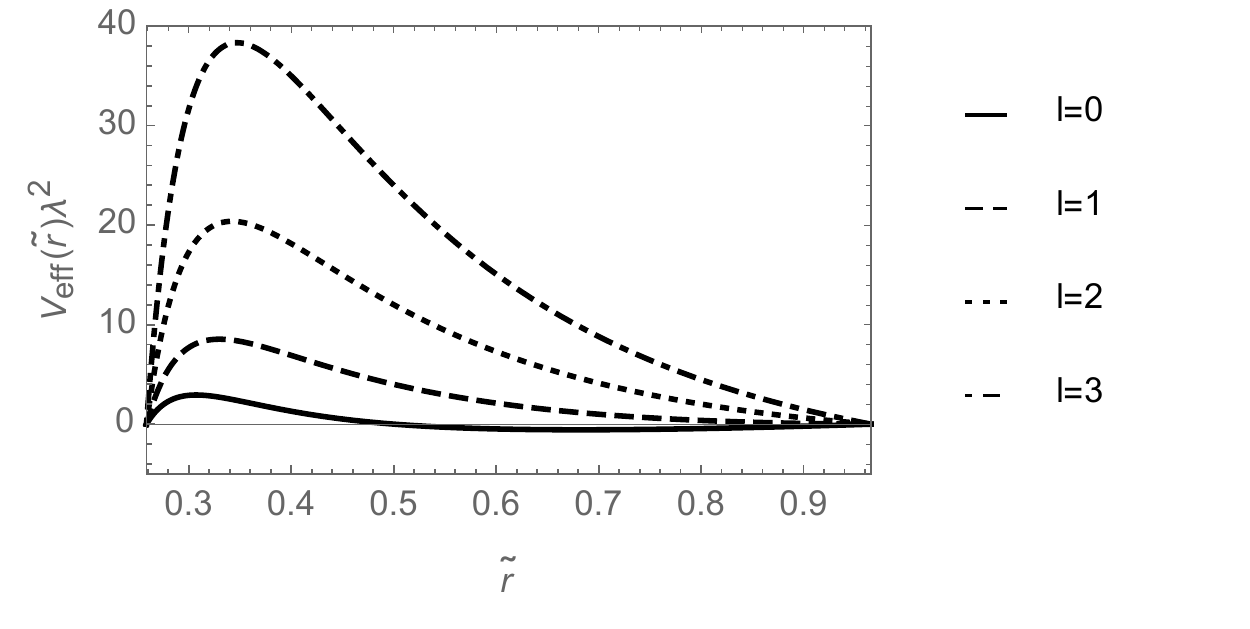}
\includegraphics[width=0.45\textwidth]{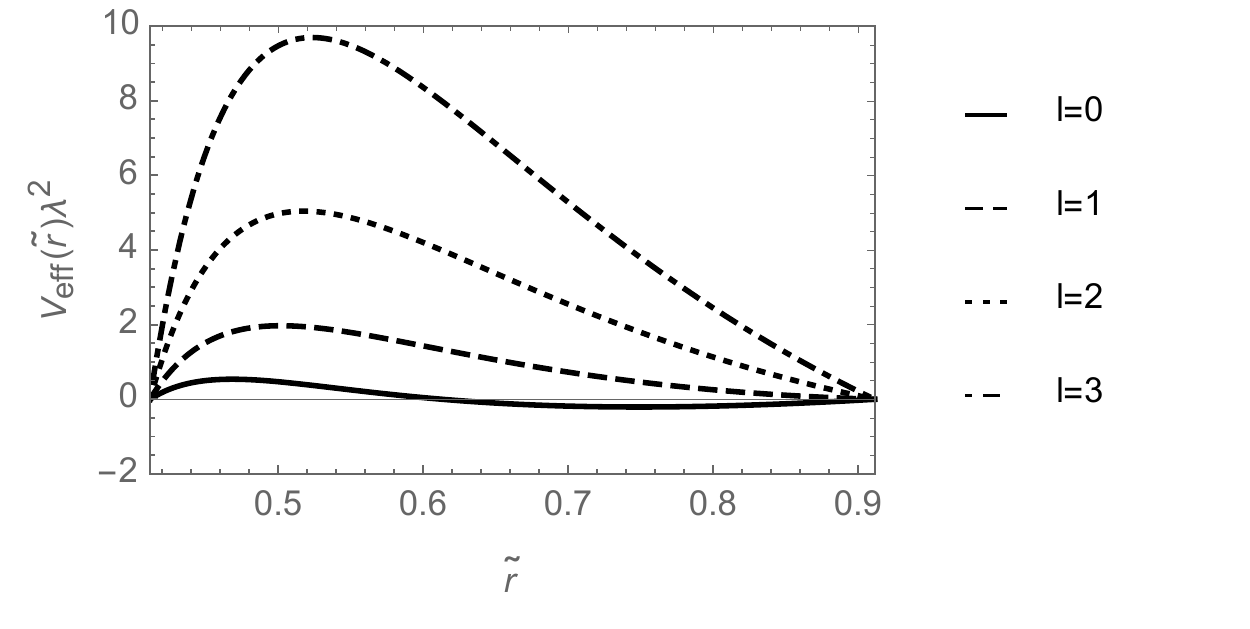}
\end{center}
\caption{The behaviour of $V_{eff}(\tilde{r})\lambda^2$ for massive scalar field with  $m=0.1$. Left panel for $\tilde{Q}=0.50$,  $\tilde{r}_h\approx 0.259$ and $\tilde{r}_c\approx 0.966$, and right panel for $\tilde{Q}=0.75$, $\tilde{r}_h\approx 0.411$ and $\tilde{r}_c\approx 0.911$.}
\label{Potential}
\end{figure}

Now, by means of the ansatz $\psi =e^{-i\omega t} Y_{l,m}(\theta,\phi) R(r)$ the Klein-Gordon equation (\ref{KGE}) can be written as 
\begin{equation}
\frac{1}{r^{2}}\frac{d}{dr}\left(r^{2} f(r)\frac{dR}{dr}\right)+\left(\frac{\omega^2}{f(r)}+\frac{\kappa^2}{r}-m^{2} \right) R(r)=0\,, \label{radial}
\end{equation}
where $\kappa^2 = -l(l+1)$, with $l = 0, 1, 2, ...$ that represent the eigenvalues of the Laplacian on the two-sphere
and $l$ is the multipole number or the angular momentum of the field. Now, defining $R(r) =\frac{F (r)}{r}$
and the tortoise coordinate $dr^{*} =\frac{dr}{f(r)}$
the wave equation can be written as a one-dimensional Schr\"odinger-like equation given by 
\begin{equation}
 \label{ggg}
 -\frac{d^{2}F(r^*)}{dr^{*2}}+V_{eff}(r)F(r^*)=\omega^{2}F(r^*)\,,
 \end{equation}
 with an effective potential $V_{eff}(r)$, which is parametrically thought as $V_{eff}(r^*)$, and it is given by
 \begin{equation}
    V_{eff}(r) =\frac{\left(-\frac{\tilde{Q}^2}{4 \tilde{r}^2}-\tilde{r}^2+1\right) \left(l^2+l+\tilde{r}^2 \left(\tilde{m}^{2}-2\right)+\frac{\tilde{Q}^2}{2 \tilde{r}^2}\right)}{\lambda^2 \tilde{r}^2}\,,
 \end{equation}
 where we have defined the dimensionless quantities $\tilde{r} \equiv r/\lambda$, $\tilde{Q} \equiv Q /\lambda$ and $\tilde{m} \equiv   \lambda m$. In Fig. \ref{Potential} we show the effective potential, for fixes values of the parameter $\tilde{Q}$, different values of the angular number $l$ and for a mass $m=0.1$ of the scalar field. It is possible to observe that for $l=0$, part of the effective potential is negative, and when $l$ increases the height of the potential barrier increases. However, when the parameter $\tilde{Q}$ increases the height of the potential barrier decreases. It is worth noting that by means of the following identification $\lambda=\sqrt{\frac{3}{\Lambda_{eff}}}$ the effective spacetime potential is asymptotically de Sitter and tends to $-\frac{r^{2}}{\lambda^{4}}(m^2\lambda^{2}-2)$ for $l=0$ reproducing the result obtained in \cite{Aragon:2020tvq}.

\newpage

\section{Quasinormal modes}
\label{QNM}

\subsection{Photon sphere modes}

{\bf{Anomalous decay rate and an approach to the critical scalar field mass.}}
In order to get some analytical insight of the behaviour 
of the QNFs, and to determine the critical scalar field mass, we use the WKB method at third order \cite{Mashhoon, Schutz:1985km, Iyer:1986np, Konoplya:2003ii, Matyjasek:2017psv, Konoplya:2019hlu}. The WKB method can be used for effective potentials which have the form of a barrier potential, approaching to a constant value at the event horizon and at the cosmological horizon or spatial infinity \cite{Konoplya:2011qq}. Here, we consider the eikonal limit $l \rightarrow \infty$ to estimate the critical scalar field mass, by
considering $\omega_I^l=\omega_I^{l+1}$ as a proxy for where the transition or critical behaviour occurs \cite{Lagos:2020oek}. The QNMs are determined by the behaviour of the effective potential near its maximum value $r^*_{max}$. The Taylor series expansion of the potential around its maximum is given by
\begin{equation}
V(r^*)= V(r^*_{max})+ \sum_{i=2}^{\infty} \frac{V^{(i)}}{i!} (r^*-r^*_{max})^{i} \,,
\end{equation}
where
\begin{equation}
V^{(i)}= \frac{d^{i}}{d r^{*i}}V(r^*)|_{r^*=r^*_{max}}\,,
\end{equation}
corresponds to the $i$-th derivative of the potential with respect to $r^*$ evaluated at the location of the maximum of the potential. Using the WKB approximation up to third order the QNFs are given by the following expression \cite{Hatsuda:2019eoj}

\begin{eqnarray}\label{frecuencie}
\omega^2 &=& V(r^*_{max})  -2 i U \,,
\end{eqnarray}
where
\begin{eqnarray}
\notag U &=&  N\sqrt{-V^{(2)}/2}+\frac{i}{64} \left( -\frac{1}{9}\frac{V^{(3)2}}{V^{(2)2}} (7+60N^2)+\frac{V^{(4)}}{V^{(2)}}(1+4 N^2) \right)  +\frac{N}{2^{3/2} 288} \Bigg( \frac{5}{24} \frac{V^{(3)4}}{(-V^{(2)})^{9/2}} (77+188N^2) + \\
\notag &&  \frac{3}{4} \frac{V^{(3)2} V^{(4)}}{(-V^{(2)})^{7/2}}(51+100N^2) +\frac{1}{8} \frac{V^{(4)2}}{(-V^{(2)})^{5/2}}(67+68 N^2)+\frac{V^{(3)}V^{(5)}}{(-V^{(2)})^{5/2}}(19+28N^2)+\frac{V^{(6)}}{(-V^{(2)})^{3/2}} (5+4N^2)  \Bigg)\,,
\end{eqnarray}
and $N=n_{PS}+1/2$, with $n_{PS}=0,1,2,\dots$, is the overtone number. Now, defining $L^2= l (l+1)$, we find that for large values of $L$, the maximum of the potential is approximately at
\begin{equation}
\frac{r_{max}}{\lambda}\approx\frac{-\sqrt{2} \tilde{m}^2 \tilde{Q}^5+\sqrt{2}  \tilde{m}^2  \tilde{Q}^3+2 \sqrt{2} \tilde{Q}^5-2 \sqrt{2}  \tilde{Q}}{16 L^2}+\frac{\tilde{Q}}{\sqrt{2}}\,,
\end{equation}
and
\begin{equation}
  \lambda^2 V(r^*_{max}) \approx \left(\frac{1}{\tilde{Q}^2}-1 \right)L^{2}+\left(-2-\frac{ \tilde{m}^2 \tilde{Q}^2}{2}+\frac{\tilde{m}^2}{2}+\tilde{Q}^2+\frac{1}{\tilde{Q}^2}\right)\,,
\end{equation}
while the second derivative of the potential evaluated at $r^*_{max}$ yields
\begin{equation}
 \lambda^4  V^{(2)}(r^*_{max}) \approx\frac{\left(1-\tilde{Q}^2\right)^2 \left(-6-3 \tilde{m}^2 \tilde{Q}^4+\tilde{m}^2 \tilde{Q}^2+6 \tilde{Q}^4\right)}{\tilde{Q}^4}-\frac{L^2 \left(4 \left(\tilde{Q}^2-1\right)^2\right)}{\tilde{Q}^4}\,.
\end{equation}
For the higher derivatives of the potential, we consider only the leading terms that are important in the limit considered. So, 
\begin{eqnarray}
  \lambda^5 V^{(3)}(r^*_{max}) && \approx  -\frac{L^2 \left(6 \sqrt{2} \left(-1+\tilde{Q}^6-3\tilde{Q}^4+3 \tilde{Q}^2\right)\right)}{\tilde{Q}^5}\,,\\
  \lambda^6V^{(4)}(r^*_{max}) && \approx  L^2 \left(-40+\frac{58}{\tilde{Q}^6}-\frac{168}{\tilde{Q}^4}-6 \tilde{Q}^2+\frac{156}{\tilde{Q}^2}\right)\,,\\
 \lambda^7 V^{(5)}(r^*_{max})  && \approx -\frac{L^2 \left(360 \sqrt{2} \left(1+\tilde{Q}^8-4 \tilde{Q}^6+6  \tilde{Q}^4-4 \tilde{Q}^2\right)\right)}{\tilde{Q}^7}\,,\\
 \lambda^8 V^{(6)}(r^*_{max})  && \approx L^2 \left(6344-\frac{472}{\tilde{Q}^8}+\frac{184}{\tilde{Q}^6}+\frac{3984}{\tilde{Q}^4}-1704 \tilde{Q}^2-\frac{8336}{\tilde{Q}^2}\right)\,.
\end{eqnarray}
Now, by using these results we find that $U$ evaluated at $r^*_{max}$ is approximately given  by
\begin{eqnarray}
 \notag \lambda^2 U\approx \frac{B\left(n_{PS}+\frac{1}{2}\right)}{\sqrt{2}}+\frac{1}{576\sqrt{2}}\left(n_{PS}+\frac{1}{2}\right)\Bigg(\frac{1080L^{8}\left(77+188\left(n_{PS}+\frac{1}{2}\right)^{2}\right)(\tilde{Q}^{2}-1)^{12}}{\tilde{Q}^{20}B^{9}}\\
 \notag 
-\frac{108 L^6 \left(100 \left(n_{PS}+\frac{1}{2}\right)^2+51\right) \left(\tilde{Q}^2-1\right)^9 \left(29+3 \tilde{Q}^2\right)}{B^7  \tilde{Q}^{16}}+\frac{4 L^4 \left(68 \left(n_{PS}+\frac{1}{2}\right)^2+67\right) \left(\tilde{Q}^2-1\right)^6 \left(29 +3 \tilde{Q}^2\right)^2}{8B^{5} \tilde{Q}^{12}}\nonumber\\
+\frac{4320 L^{4} \left(28\left(n_{PS}+\frac{1}{2}\right)^2+19\right) \left(\tilde{Q}^{2}-1\right)^7}{B^{5}  \tilde{Q}^{12}}-\frac{L^2 \left(4 \left(n_{PS}+\frac{1}{2}\right)^{2}+5\right) \left(\tilde{Q}^{2}-1\right)^{4} \left(59 +213 \tilde{Q}^{2}\right)}{B^{3}  \tilde{Q}^{8}}\Bigg)\nonumber\\
+\frac{i}{64}\left(-\frac{8 L^{4}\left(60\left(n_{PS}+\frac{1}{2}\right)^{2}+7\right) \left(\tilde{Q}^{2}-1\right)^{6}}{B^4 
 \tilde{Q}^{10}}+\frac{2 L^{2} \left(4\left(n_{PS}+\frac{1}{2}\right)^{2}+1\right) \left(\tilde{Q}^{2}-1\right)^{3} \left(29 +3 \tilde{Q}^{2}\right)}{B^2 \tilde{Q}^{6}}\right)\,,
\end{eqnarray}
where $B \equiv \frac{\left(\tilde{Q}^2-1\right)\sqrt{\left(2 \left(2 L^2+3\right)+3 \tilde{Q}^4 \left(\tilde{m}^2-2\right)-\tilde{m}^2 \tilde{Q}^2\right)}}{\tilde{Q}^2}$.
So, using these results together with Eq. (\ref{frecuencie})   we obtain
the following analytical QNFs that is valid for large values of $L$:
\begin{equation}\label{frequency}
\tilde{\omega} \equiv \lambda \omega 
 \approx 
 \tilde{\omega}_{1m}L+\tilde{\omega}_{0}+\tilde{\omega}_{1}L^{-1}+\tilde{\omega}_{2}L^{-2}\,,
\end{equation}
where
\begin{equation}
\tilde{\omega}_{1m}=\frac{\sqrt{1-\tilde{Q}^{2}}}{\tilde{Q}}\,,\,\,
\tilde{\omega}_{0}=\frac{i\left(-1-2n_{PS}+2n_{PS} \tilde{Q}^{2}+ \tilde{Q}^{2}\right)}{\tilde{Q}\sqrt{2}\sqrt{1-\tilde{Q}^2}}\,,
\end{equation}
\begin{equation}
\tilde{\omega}_{1}=\frac{\sqrt{1-\tilde{Q}^2} \left(\left(2 \tilde{m}^2 \tilde{Q}^2-3 n_{PS} (n_{PS}+1)+1\right)+3 \left(n_{PS}(n_{PS}+1) -1\right) \tilde{Q}^2\right)}{8 \tilde{Q}}\,,
\end{equation}
\begin{equation}
\tilde{\omega}_{2}=\frac{i (2 n_{PS}+1) \left(1-\tilde{Q}^2\right)^{3/2} \left(\tilde{Q}^2 \left(48 \tilde{m}^2+17 n_{PS} (n_{PS}+1)-87\right)- (17 n_{PS} (n_{PS}+1)+9)\right)}{128 \sqrt{2} \tilde{Q}}\,.
\end{equation}
Now, the term proportional to $1/L^{2}$ is zero at the value of the critical mass $\tilde{m}_{c}$, which is given by
\begin{equation}
\label{eqmass}
\tilde{m}_c \equiv \lambda m_c = \frac{\sqrt{17 n_{PS} (n_{PS}+1)+9+(87-17 n_{PS} (n_{PS}+1)) \tilde{Q}^2}}{4 \sqrt{3} \tilde{Q}}\,.
\end{equation}

\begin{figure}[h]
\begin{center}
\includegraphics[width=0.32\textwidth]{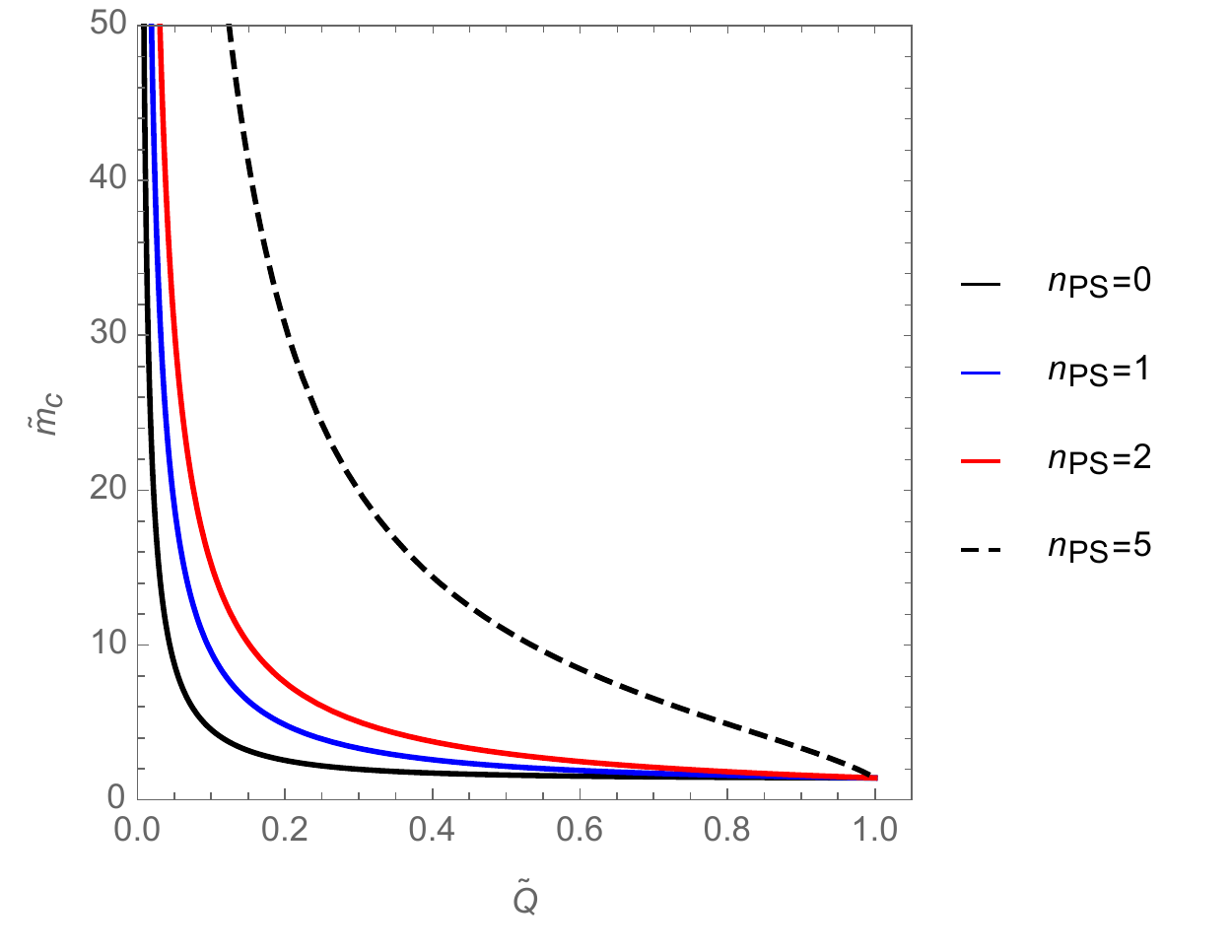}
\end{center}
\caption{The behaviour of $\tilde{m}_c$ as a function of $\tilde{Q}$ for different values of the overtone number $n_{PS}=0,1,2$ and $5$.} 
\label{FMC}
\end{figure}

In Fig. \ref{FMC}, we show the behaviour of $\tilde{m}_c$ as a function of $\tilde{Q}$. We observe that $\tilde{m}_c$ decreases when $\tilde{Q}$ increases, and for a fixed value of $\tilde{Q}<1$, $\tilde{m}_c$ increases when the overtone number $n_{PS}$ increases. However, when $\tilde{Q} \rightarrow 1$ (or $r_c$ $\rightarrow$ $r_h$) then $\tilde{m}_c \rightarrow \sqrt{2}$, and it does not depend on the overtone number $n_{PS}$. Also, when $\tilde{Q} \rightarrow 0$, $\tilde{m}_c \rightarrow \infty$.\\

\begin{figure}[H]
\begin{center}
\includegraphics[width=0.3\textwidth]{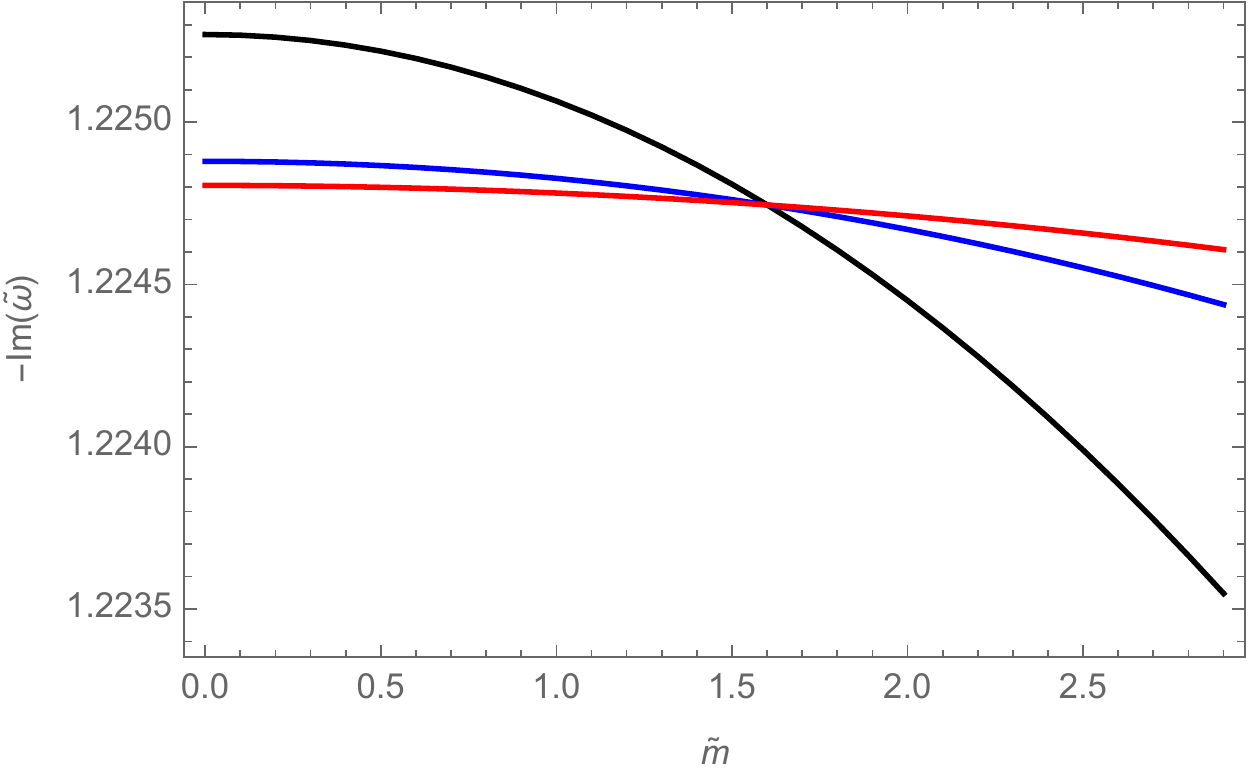}
\includegraphics[width=0.3\textwidth]{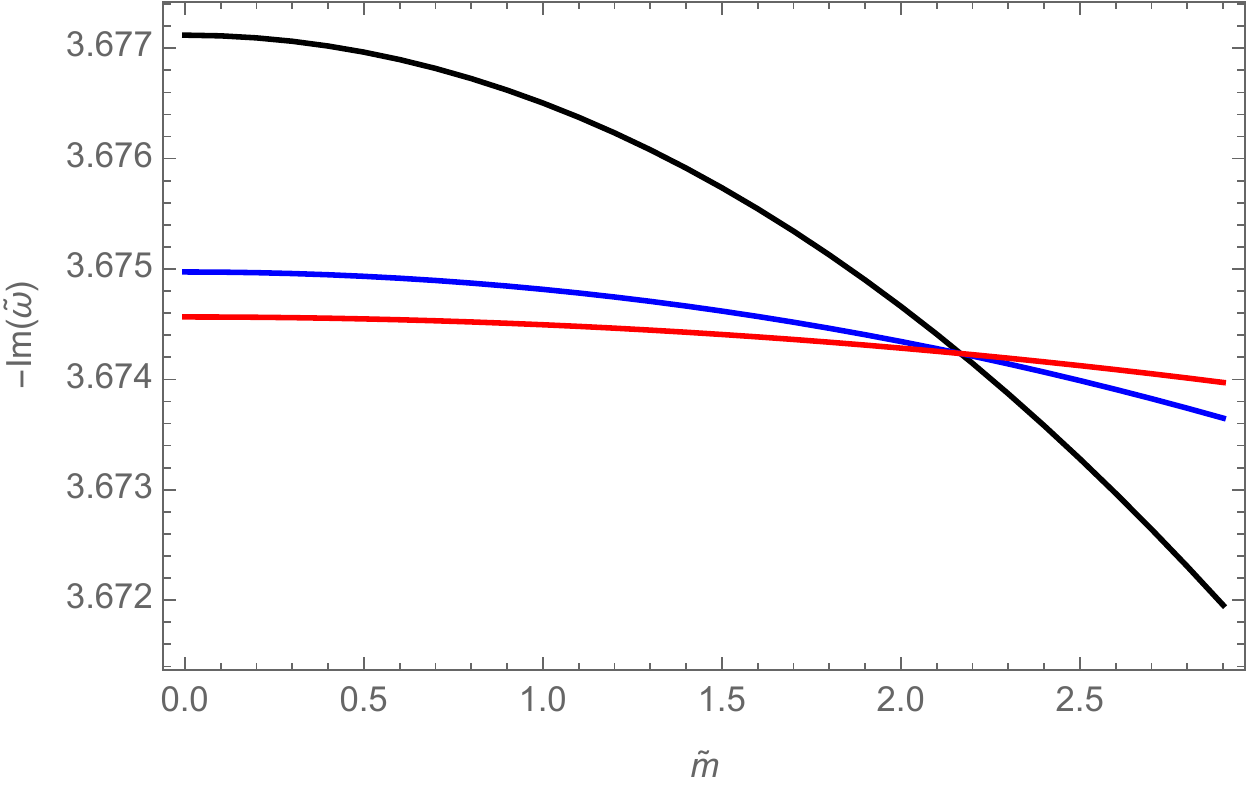}
\includegraphics[width=0.3\textwidth]{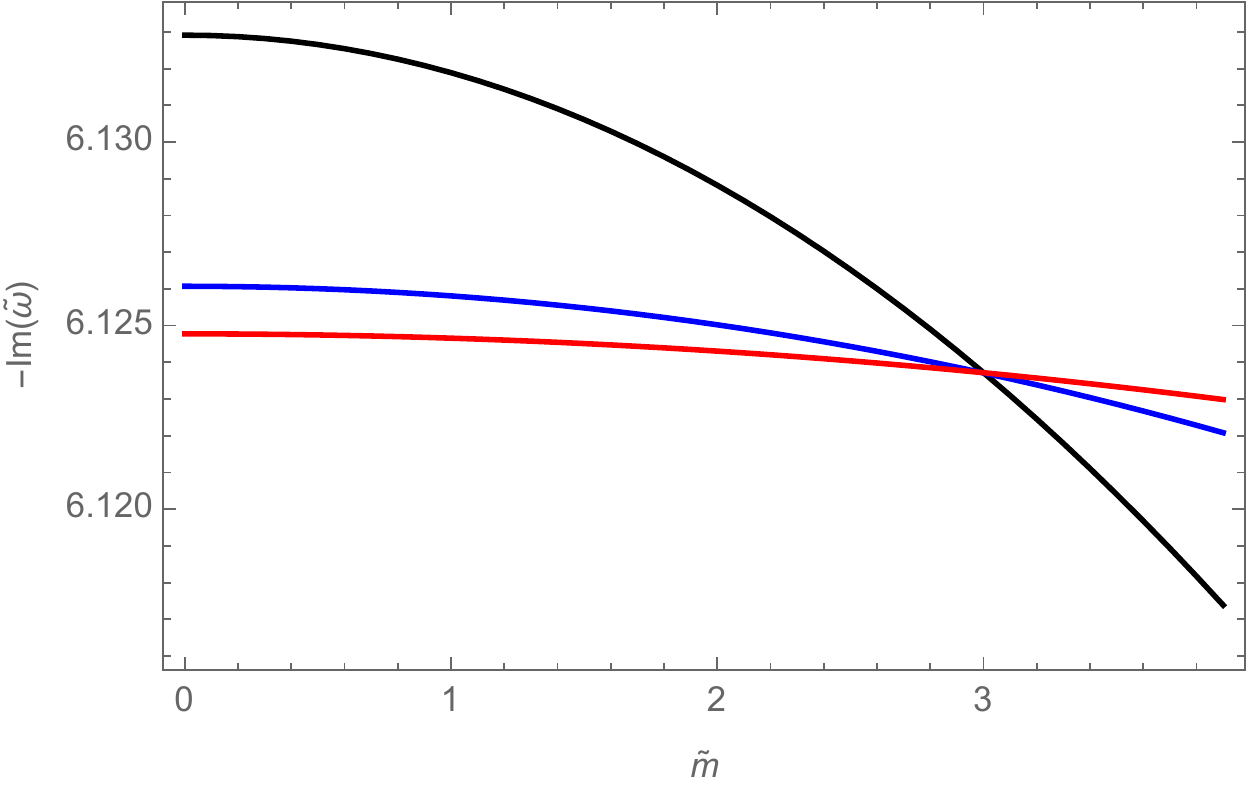}
\end{center}
\caption{
The behaviour of
$-Im(\tilde{\omega})$ as a function of  $\tilde{m}$, with $\tilde{Q}=0.5$. Left panel for $n_{PS}=0$, central panel for $n_{PS}=1$, and right panel for $n_{PS}=2$.
Here, the WKB method gives via Eq. (\ref{eqmass}) $\tilde{m}_{c} \approx 1.601$ for $n_{PS}=0$, $\tilde{m}_{c} \approx 2.165$ for $n_{PS}=1$, and $\tilde{m}_{c} \approx 2.990$ for $n_{PS}=2$. Black lines for $l=20$, blue lines for $l=40$, and red lines for $l=60$.}
\label{FA1}
\end{figure}

\begin{figure}[H]
\begin{center}
\includegraphics[width=0.3\textwidth]{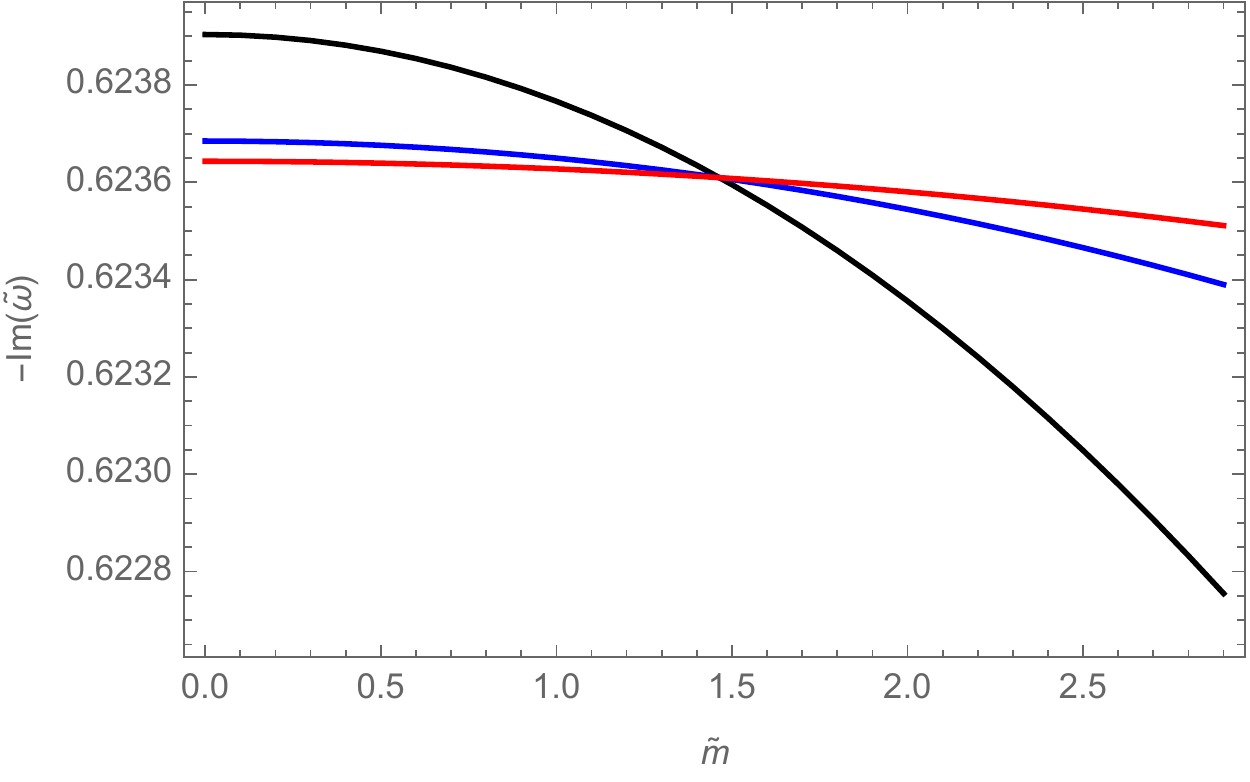}
\includegraphics[width=0.3\textwidth]{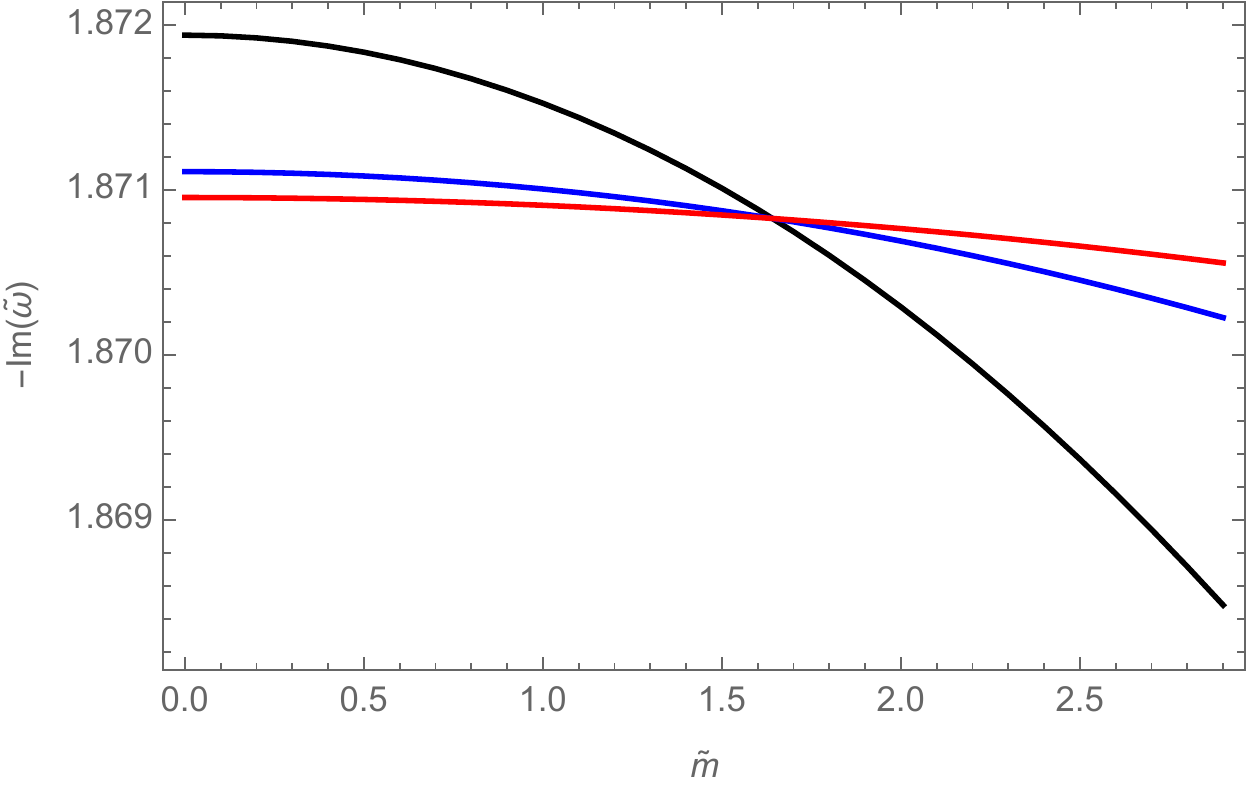}
\includegraphics[width=0.3\textwidth]{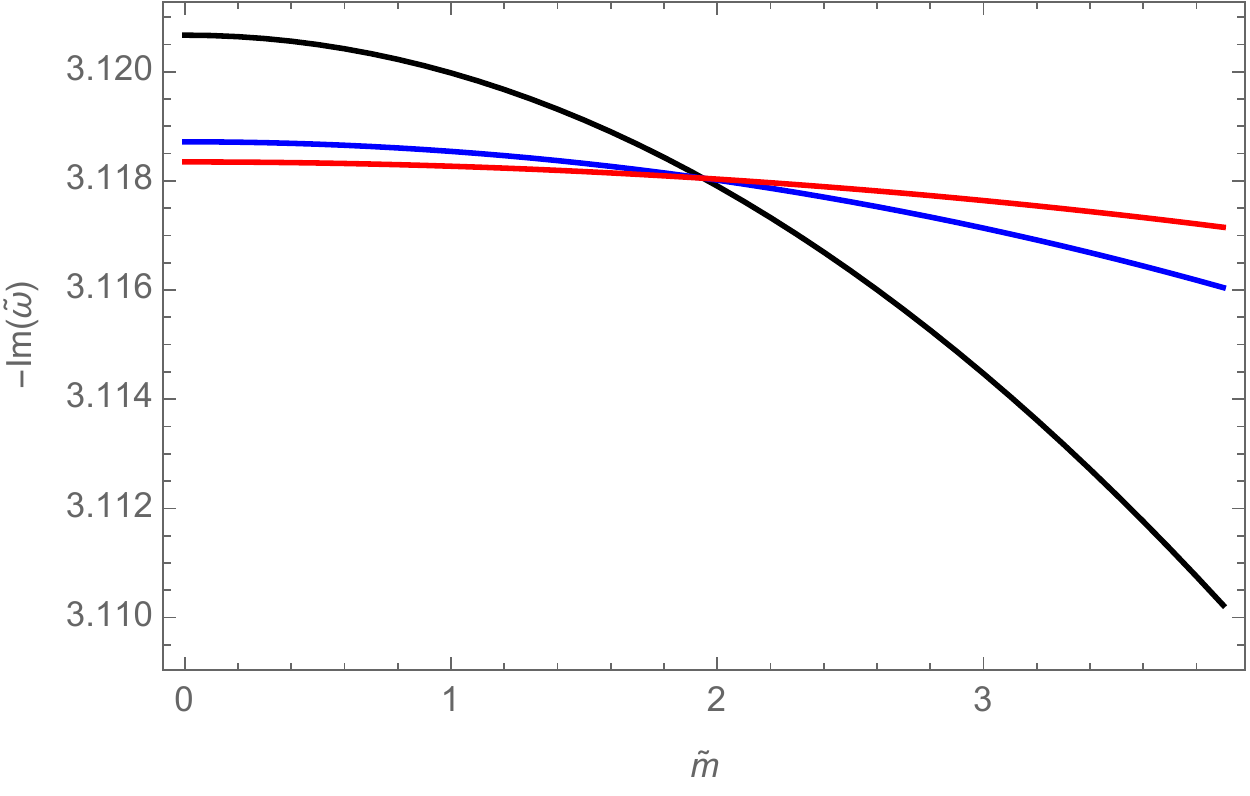}
\end{center}
\caption{
The behaviour of
$-Im(\tilde{\omega})$ as a function of  $\tilde{m}$, with $\tilde{Q}=0.75$. Left panel for $n_{PS}=0$, central panel for $n_{PS}=1$, and right panel for $n_{PS}=2$.
Here, the WKB method gives via Eq. (\ref{eqmass}) $\tilde{m}_{c} \approx 1.465$ for $n_{PS}=0$, $\tilde{m}_{c} \approx 1.642$ for $n_{PS}=1$, and $\tilde{m}_{c} \approx 1.949$ for $n_{PS}=2$. Black lines for $l=20$, blue lines for $l=40$, and red lines for $l=60$.}
\label{FA2}
\end{figure}

{\bf{Anomalous decay rate.}} Here, we use the 6th order WKB  method in order to show the anomalous decay rate by simplicity; however, at the end we compare the QNFs via the 6th order WKB  method and the the pseudospectral Chebyshev method to show the accuracy of the 6th order WKB  method. So, in Figs. \ref{FA1}, and \ref{FA2}, we show the behaviour of $-Im(\tilde{\omega})$ as a function of $\tilde{m}$. We can observe an anomalous decay rate, i.e,  for $\tilde{m}<\tilde{m}_c$, the long-livest modes are the one with highest angular number $l$; whereas, for $\tilde{m}>\tilde{m}_c$, the long-livest modes are the one with smallest angular number. Also, when the parameter $\tilde{Q}$ increases the parameter $\tilde{m}_c$ decreases, and when the overtone number $n_{PS}$ increases the parameter $\tilde{m}_c$ increases.

\begin{figure}[H]
\begin{center}
\includegraphics[width=0.3\textwidth]{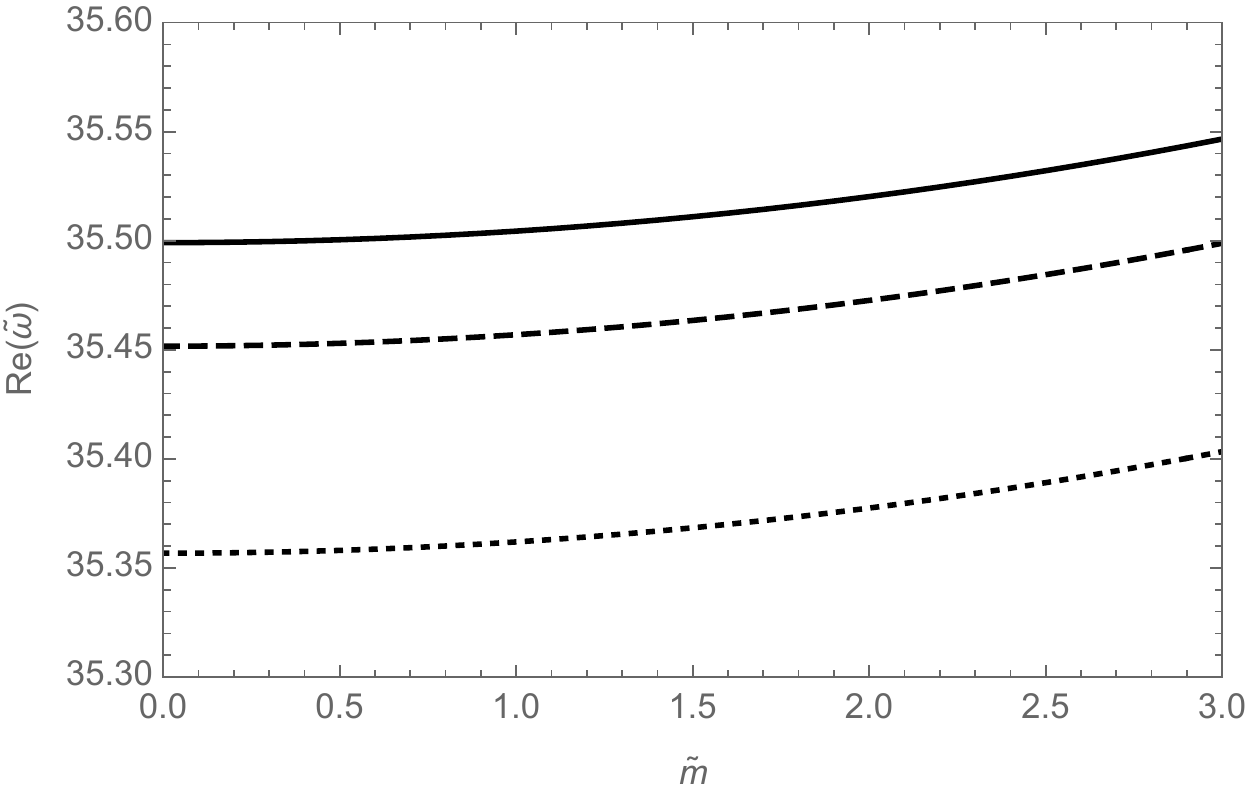}
\includegraphics[width=0.3\textwidth]{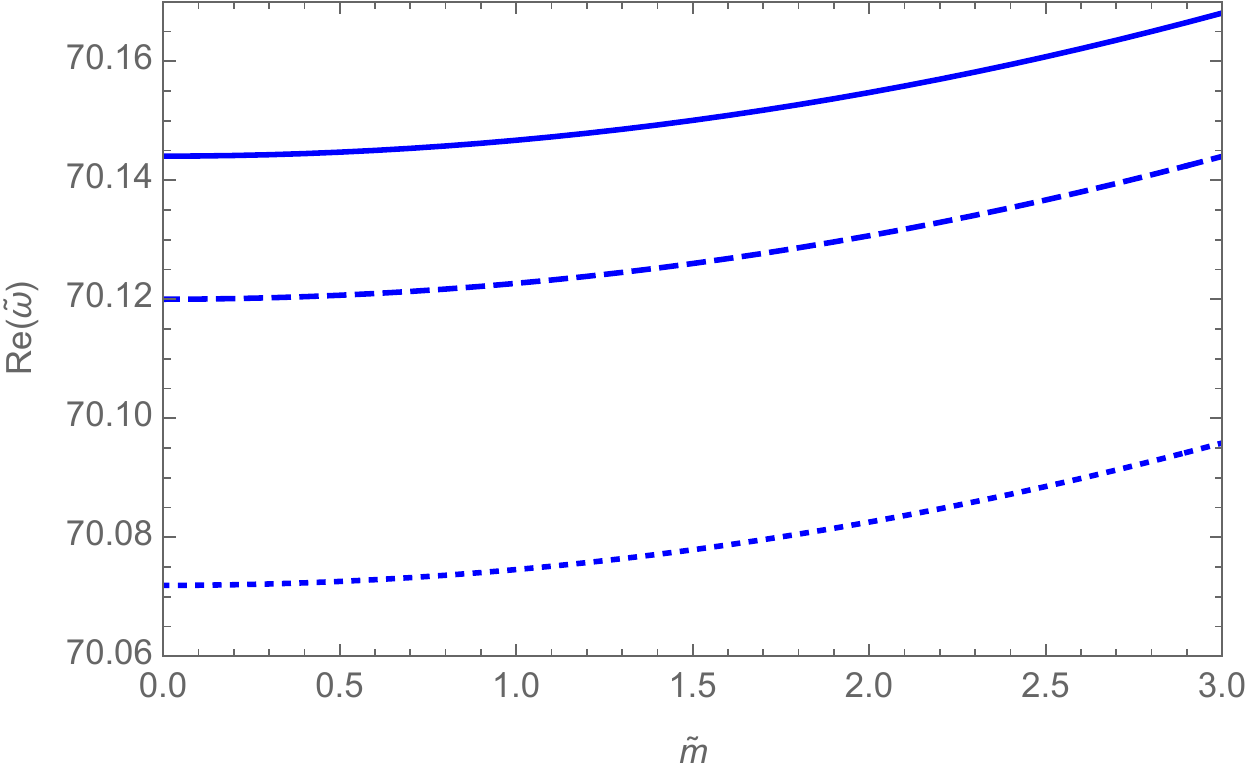}
\includegraphics[width=0.3\textwidth]{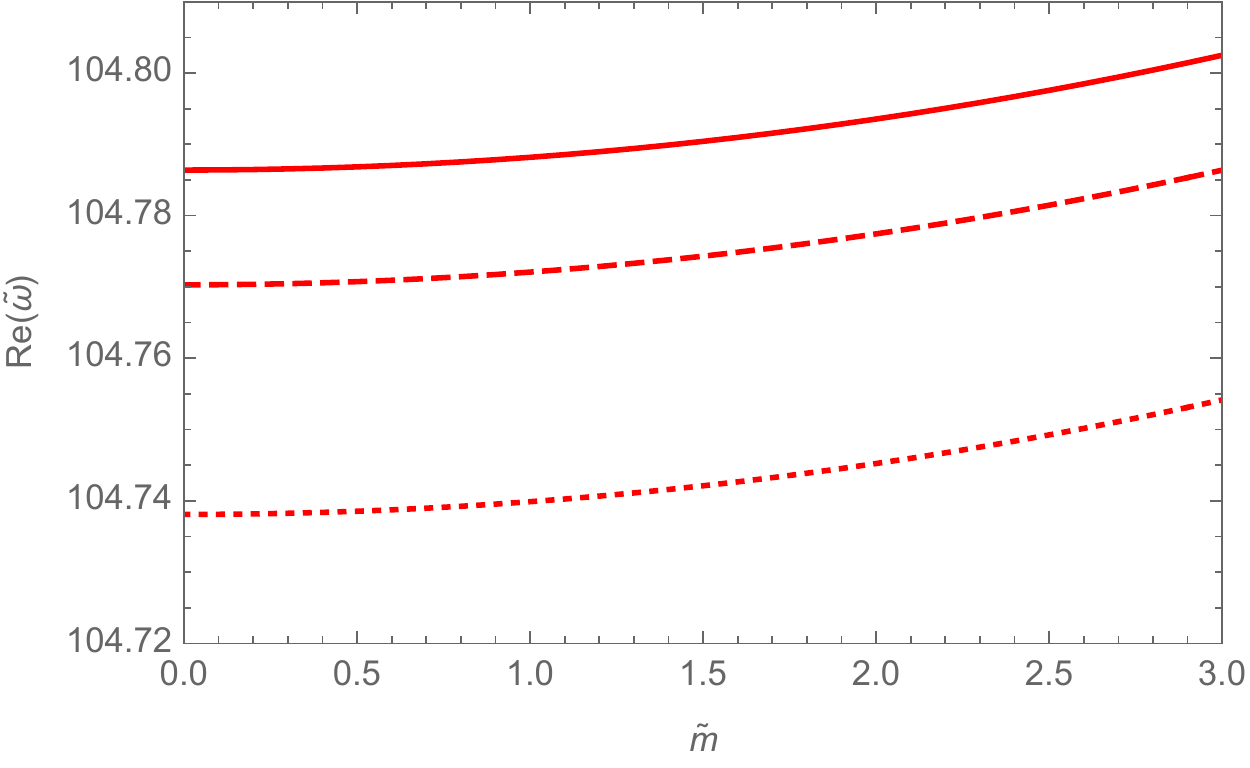}
\end{center}
\caption{
The behaviour of
$Re(\tilde{\omega})$ as a function of  $\tilde{m}$, with $\tilde{Q}=0.5$. Left panel for $l=20$, central panel for $l=40$, and right panel for $l=60$.
Solid lines for $n_{PS}=0$, dashed lines for $n_{PS}=1$, and dotted lines for $n_{PS}=2$.}
\label{RAA1}
\end{figure}

\begin{figure}[H]
\begin{center}
\includegraphics[width=0.3\textwidth]{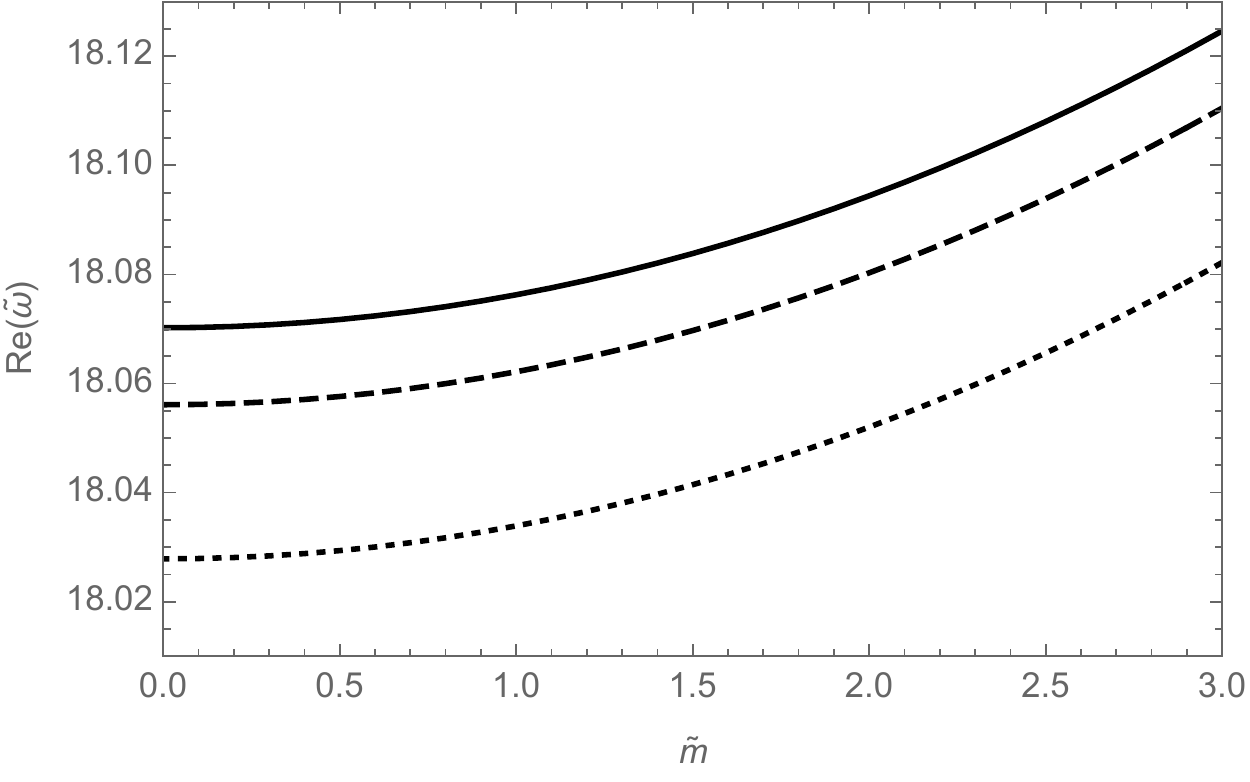}
\includegraphics[width=0.3\textwidth]{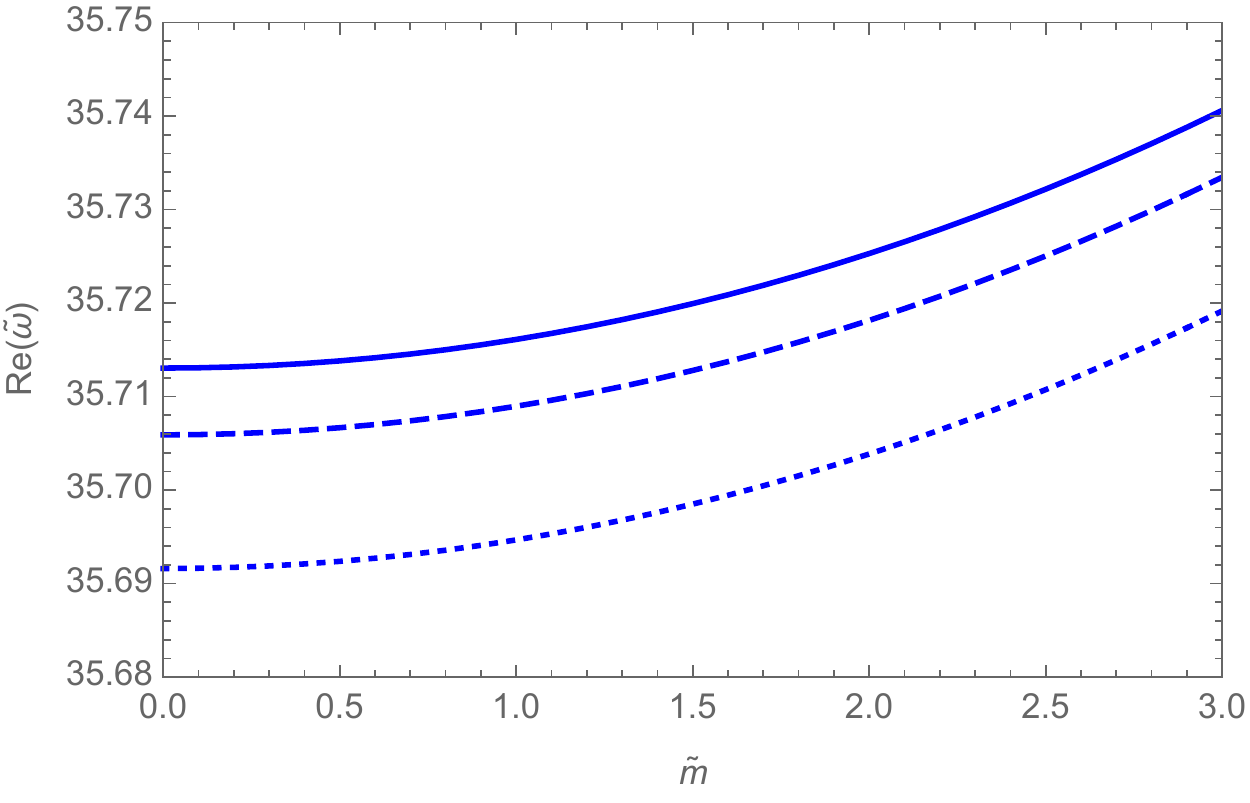}
\includegraphics[width=0.3\textwidth]{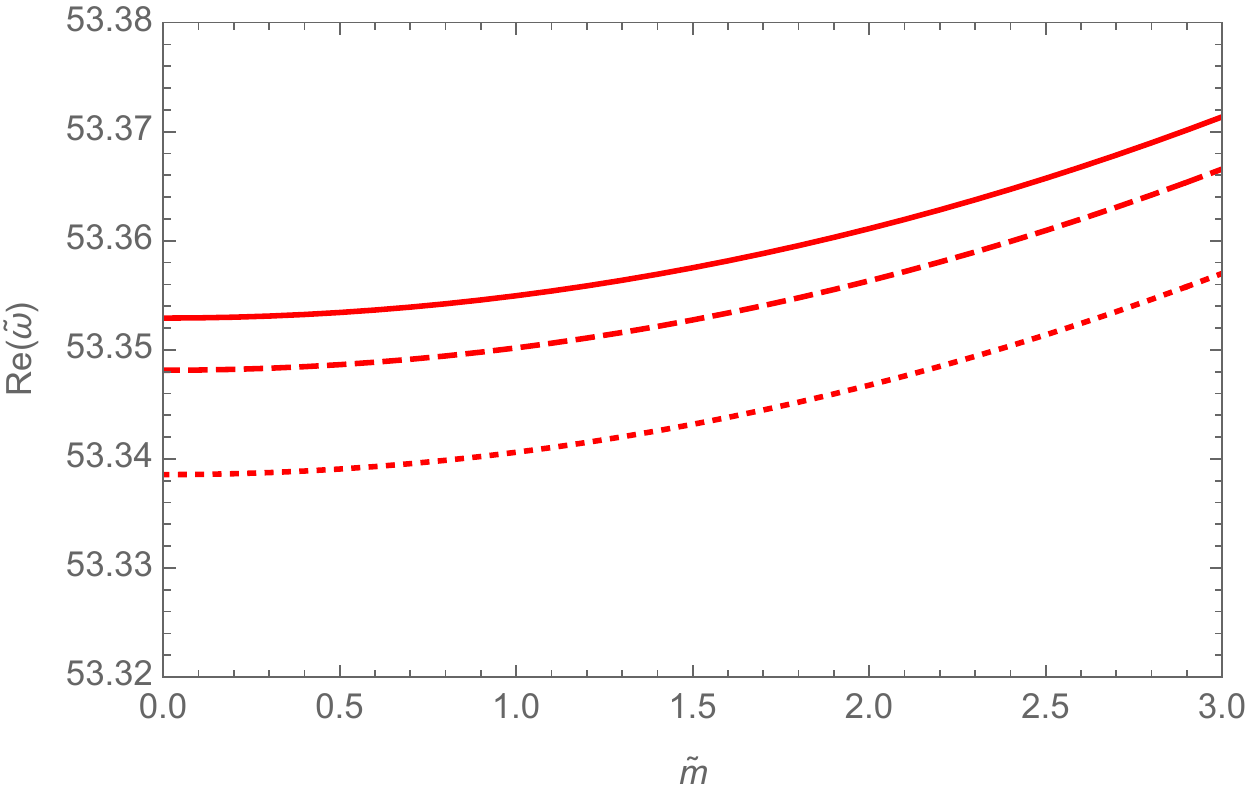}
\end{center}
\caption{
The behaviour of
$Re(\tilde{\omega})$ as a function of  $\tilde{m}$, with $\tilde{Q}=0.75$. Left panel for $l=20$, central panel for $l=40$, and right panel for $l=60$.
Solid lines for $n_{PS}=0$, dashed lines for $n_{PS}=1$, and dotted lines for $n_{PS}=2$.}
\label{RAA2}
\end{figure}

Now, in Fig. \ref{RAA1}, and Fig. \ref{RAA2}, we show the behaviour of $Re(\tilde{\omega})$ as a function of  $\tilde{m}$, we can observe that the frequency of oscillation increases when the scalar field mass increases, and the frequency of oscillation decreases when the overtone number increases. Also, when the parameter $\tilde{Q}$ increases the frequency of oscillation decreases.

{\bf{Accuracy of the numerical techniques.}} Now, in order to check the correctness and accuracy of the 6th order WKB method, we will compare some QNFs with the pseudospectral Chebyshev method. Thus, we show in Table \ref{QNMC}, see appendix \ref{tables}, the QNFs by using the pseudospectral Chebyshev method and the WKB method. Also, we show the relative error, which is defined by
\begin{eqnarray}
\label{ERe}
\epsilon_{Re(\tilde{\omega})} =\frac{\mid Re(\tilde{\omega}_1)\mid- \mid Re(\tilde{\omega}_0)\mid}{\mid Re(\tilde{\omega}_0)\mid}\cdot 100\%\,,\,\, \text{and}\,\, \epsilon_{Im(\tilde{\omega})} =\frac{\mid Im(\tilde{\omega}_1)\mid- \mid Im(\tilde{\omega}_0)\mid}{\mid Im(\tilde{\omega}_0)\mid} \cdot 100\%\,,
\end{eqnarray}
%\begin{equation}
%\label{EIm}
%\end{equation}
where $\tilde{\omega}_1$ corresponds to the QNFs via the 6th order WKB method, and $\tilde{\omega}_0$ denotes the QNFs via the pseudospectral Chebyshev method. We can observed that the error does not exceed $117.964$ $(\%)$ in the imaginary part, and $40.308$ $(\%)$ in the real part. These maximum values  occur for small values of $l$, where the 6th order WKB method does not provide a high accuracy, it is known that the WKB method provides better accuracy for larger $l$ (and $l>n$). Note that the error increases for higher values of the scalar field mass and for higher values of the overtone number. Note also that the WKB method has a good accuracy, for $l\geq 20$, where the error does not exceed $8.813\cdot 10^{-6}$ $(\%)$ in the imaginary part, and $1.639\cdot 10^{-7}$ $(\%)$ in the real part. So, this method with $l\geq 20$ is appropriate in order to show the anomalous behaviour of the decay rate.

\subsection{de Sitter modes}

These modes are associated with the presence of the cosmological horizon,  in this spacetime the effective cosmological constant $\Lambda_{eff} = 3/ \lambda^2$ provides the asymptotically de Sitter solution,
and resemble those of a pure de Sitter spacetime \cite{Du:2004jt}, which are given by
\begin{equation}\label{deS}
\omega_{pure-dS} = - i \sqrt{\frac{\Lambda}{3}} \left( 2n_{dS} + l + 3/2 \pm \sqrt{\frac{9}{4}-3 \frac{m^2}{\Lambda}} \right)\,,
\end{equation} 
where $\Lambda$ is the (positive) cosmological constant of pure de Sitter spacetime. It it worth to notice that for $m^2 \leq 3 \Lambda/4$ the QNFs of pure de Sitter spacetime are purely imaginary whereas for $m^2 > 3 \Lambda/4$ the QNFs acquire a real part. We can observe in Table \ref{QNMell} that %for small values of the effective cosmological constant, i.e, 
for $Q=1$ and for large values of the parameter $\lambda$ (or small values of $\tilde{Q}= Q/ \lambda$) the modes resemble those of the pure de Sitter spacetime (\ref{deS}). Also, note that for massless scalar field with $n_{dS}=0$, $l=0$, there is a branch where $\omega_{dS}=0$, the zero mode. By considering this fact one could say that the zero mode is associated with the dS family. So, in the following we will consider the zero mode as a dS mode.

\begin {table}[ht]
\caption {Quasinormal frequencies $\omega$ for massless scalar fields in the background of Weyl black holes, with $Q=1$, $l=1$, and $\lambda= 10, 40, 100$, and $200$. Here, the QNFs are obtained via the pseudospectral Chebyshev method using a number of Chebyshev polynomials in the range 95-100, the values inside the quotation marks ``...'', means that the QNF converges for a number of polynomials in the range 165-170, and ... means that there is not convergence until 170 polynomials with nine decimals places of accuracy for the QNF. The values between parenthesis are obtained via Eq. (\ref{deS}).}   
\label {QNMell}\centering
%\resizebox{10cm}{!} {
\begin {tabular} { | c |c |c |c |c |}
\hline
$n_{dS}$ &  $\lambda=10\,\,(\omega_{dS})$  & $\lambda=40\,\,(\omega_{dS})$ & $\lambda=100\,\,(\omega_{dS})$ & $\lambda=200\,\,(\omega_{dS})$ \\\hline
$0$ &
$-0.099958442 i\,\, (-0.1 i)$  & $-0.024999349 i\,\, (-0.025 i)$ & $-0.009999958 i \,\, (-0.01 i)$  & $-0.004999995 i \,\, (-0.005 i)$
\\\hline
$1$ &
$-0.300271179 i\,\, (-0.3 i)$  & $"-0.075004293 i \,\, (-0.075 i)"$ & $"-0.030000275 i \,\, (-0.03 i)"$  & ... $(-0.015 i)$
   \\\hline
$2$ &
$-0.400814152 i\,\, (-0.5 i)$  & $"-0.100013002 i \,\, (-0.125 i)"$ & 
... $  (-0.05 i)$  & ... $(-0.025 i)$
 \\\hline
$3$ &
$-0.501600806 i\,\, (-0.7 i)$  & ... $ (-0.175 i)$
&
... $ (-0.07 i)$ &  ... $(-0.035 i)$ \\\hline
\end {tabular}
\end {table}

\begin{figure}[H]
\begin{center}
\includegraphics[width=0.3\textwidth]{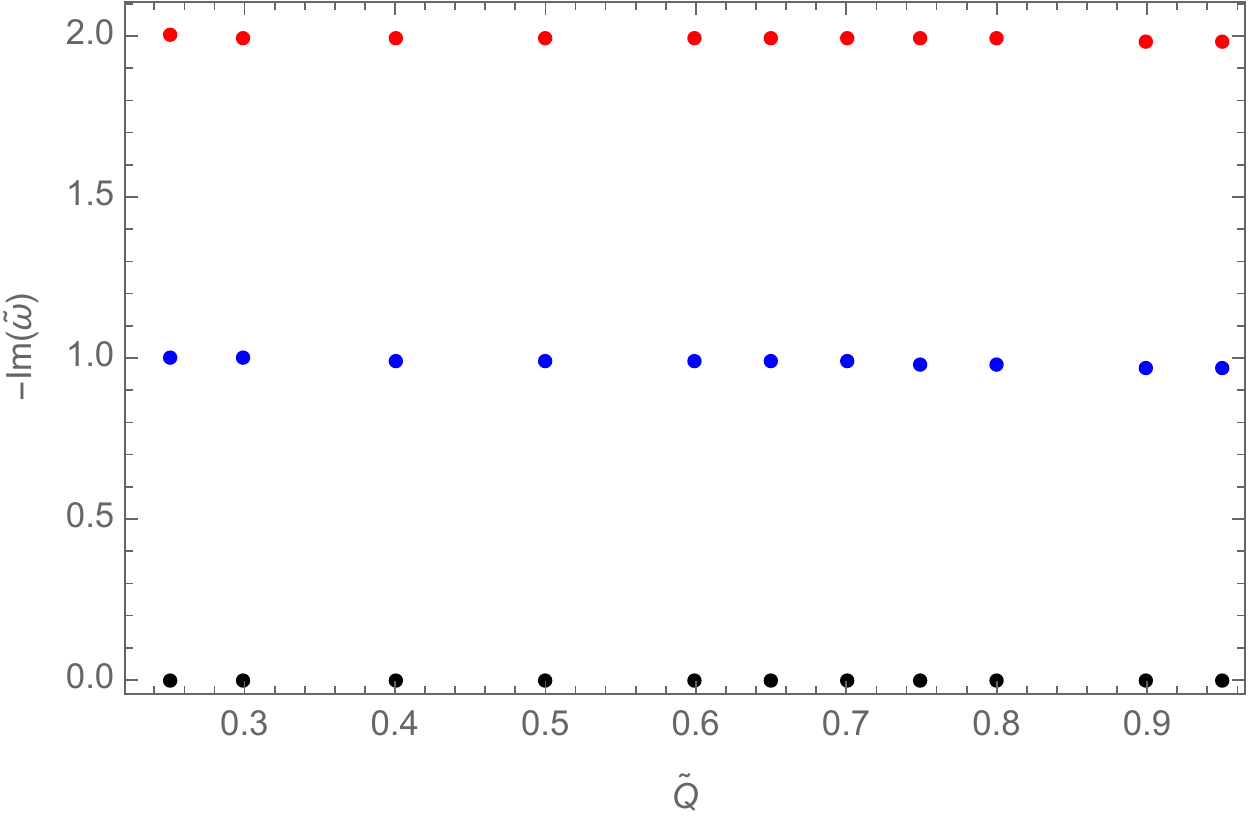}
\includegraphics[width=0.3\textwidth]{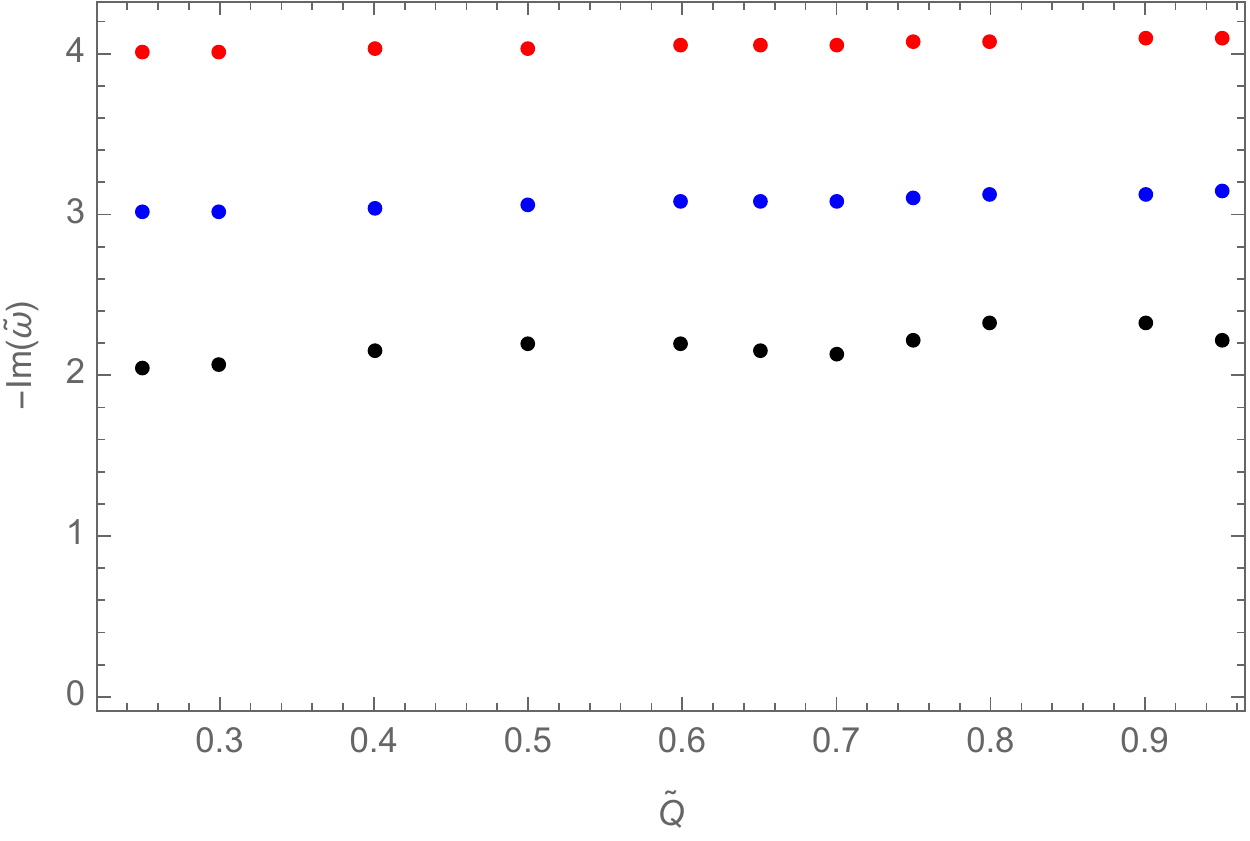}
\includegraphics[width=0.3\textwidth]{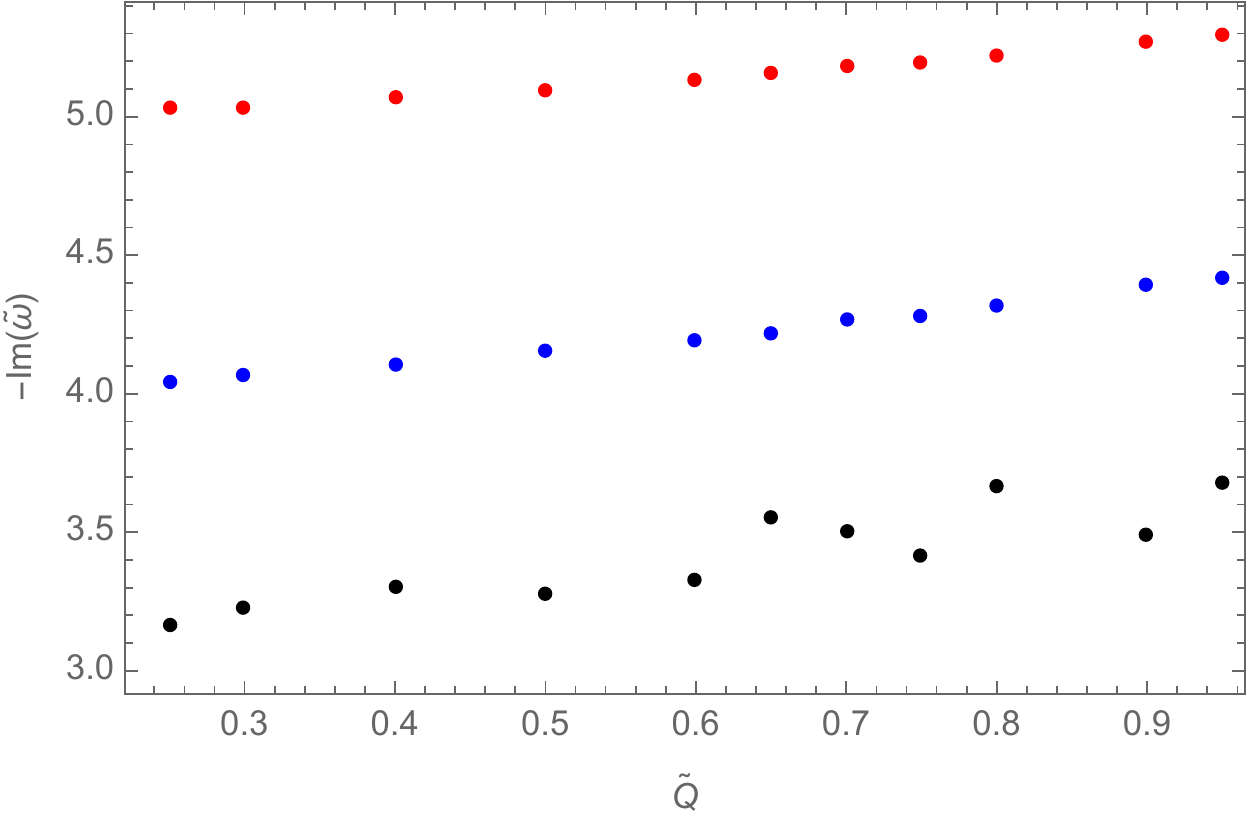}
\end{center}
\caption{
de Sitter modes $\tilde{\omega}_{dS}$ for massless scalar fields in the background of Weyl black holes for several values of $\tilde{Q}$ in the range 0.25-0.95. Black points for $l=0$, blue points for $l=1$, and red points for $l=2$. Left panel for $n_{dS}=0$, central panel for $n_{dS}=1$, and right panel for $n_{dS}=2$. Some numerical values are shown in appendix \ref{tables} Table \ref{dSmodes} .}
\label{RA1}
\end{figure}

In Fig. \ref{RA1}, we plot the behaviour of the decay rate as a function of the parameter $\tilde{Q}$, for massless scalar fields, we can observe that for {\bf{$n_{dS}=0$}}, left panel, the decay rate is not sensitive to the increment of $\tilde{Q}$, giving a null slope approximately. However, it is possible to observe a positive slope for {\bf{$n_{dS}>0$}}, and $l>0$, see central and right panels.

Now, in order to analyze the effect of the scalar field mass on the decay rate, we consider $\tilde{Q}=0.50$ in Table \ref{dS1}, and $\tilde{m}=0, 1, 1.5, 1.6, 1.7, 1.8$. We can observe that for $l=0,1,2$, $n_{dS}=0$, and purely imaginary QNFs, the decay rate increases when the scalar field mass increases. However, in general this is not true for higher overtone numbers. Also, note that the dS modes also can acquire a real part if the mass of the scalar field increases enough, which is similar to what happens to the modes of pure de Sitter spacetime.

\begin{table}[H]
\caption {de Sitter modes $\tilde{\omega}_{dS}$ for massive scalar fields in the background of Weyl black holes, with $\tilde{Q}= 0.5$. Here, the QNFs are obtained via the pseudospectral Chebyshev method using a number of Chebyshev polynomials in the range $50$-$60$, 
%and the values with $"..."$, means that the QNFs converges for iterations in the range 165-170, 
with eight decimals places of accuracy for the QNFs.}
\label {dS1}\centering
\scalebox{0.7} {

\begin {tabular} { | c | c | c | c | c | c | c | }
\hline
${}$ & $\tilde{m} = 0$ & $ \tilde{m} = 1.0$  & $ \tilde{m} = 1.5$  
&  $ \tilde{m} = 1.6$  & $ \tilde{m} = 1.7$ & $ \tilde{m} = 1.8$
\\\hline
$\tilde{\omega}_{dS} (l=0;n_{dS}=0)$ &
$0$ &
$-0.354019199 i$ &
$\pm 0.13490752-1.74387624 i$ &
$\pm 0.63744315 - 1.87301312 i$ &
$\pm 0.83471105-1.95247785 i$ &
$\pm 0.97729808 - 2.00101119 i$
 \\\hline
$\tilde{\omega}_{dS} (l=0;n_{dS}=1)$ &
$-2.20642433 i$  &
$-2.53697632 i$  &
$-3.63647842 i$  &
$\pm 0.51729099 - 3.72212001 i$ &
$\pm 0.72855748-3.68708272 i$ &
$\pm 0.88681383 - 3.66284083 i$
\\\hline
$\tilde{\omega}_{dS} (l=0;n_{dS}=2)$ &
$-3.28026122 i$ &
$-3.02091679 i$ &
$-3.92292329 i$  &
$\pm 0.41613702 - 6.35169327 i$ &
$\pm 0.64304481-6.35694024 i$ &
$\pm 0.81115469 - 6.36092768 i$

\\\hline\hline

$\tilde{\omega}_{dS} (l=1;n_{dS}=0)$ &
$-0.990611270 i$ &
$-1.379852863 i$  &
$\pm 0.02364769-2.53655476 i$  &
$\pm 0.58531081-2.52886099 i$ &
$\pm 0.83815065-2.52054994 i$ &
$\pm 1.03865944-2.51157608 i$
\\\hline

$\tilde{\omega}_{dS} (l=1;n_{dS}=1)$ &
$-3.05278804 i$  &
$-3.49374404 i$ &
$-4.67287107 i$  &
$\pm 0.61015915-4.75678420 i$ &
$\pm 0.87673848-4.77398775 i$ &
$\pm 1.08585237-4.79494463 i$
\\\hline

$\tilde{\omega}_{dS} (l=1;n_{dS}=2)$ &
$-4.15918692 i$ &
$-3.72467809 i$ &
$-4.81305355 i$  &
$\pm 0.72130136-6.96276090 i$ &
$\pm 1.03254340-6.89072258 i$ &
$\pm 1.24247629-6.80117766 i$
\\\hline \hline

$\tilde{\omega}_{dS} (l=2;n_{dS}=0)$ &
$-1.99595932 i$  &
$-2.38048725 i$  &
$-3.51812090 i$   &
$\pm 0.57636174-3.52176056 i$ &
$\pm 0.82828711-3.51855388 i$ &
$\pm 1.03034646 - 3.51518953 i$
\\\hline

$\tilde{\omega}_{dS} (l=2;n_{dS}=1)$ &
$-4.03347941 i$ &
$-4.44262643 i$ &
$-3.53147971 i$   &
$\pm 0.60772032-5.65005367 i$ &
$\pm 0.87324050-5.64738635 i$ &
$\pm 1.08633031 - 5.64432989 i$
\\\hline

$\tilde{\omega}_{dS} (l=2;n_{dS}=2)$ &
$-5.09907861 i$  &
$-4.69417428 i$  &
$\pm 0.01894009-5.65239540 i$  &
$\pm 0.62169812-7.86708187 i$ &
$\pm 0.89266625-7.86631108 i$ &
$\pm 1.10872373 - 7.86579727 i$
\\\hline

\end {tabular}
}
\end{table}\leavevmode\newline

\subsection{Dominance family modes}

\begin{figure}[H]
\begin{center}
\includegraphics[width=0.32\textwidth]{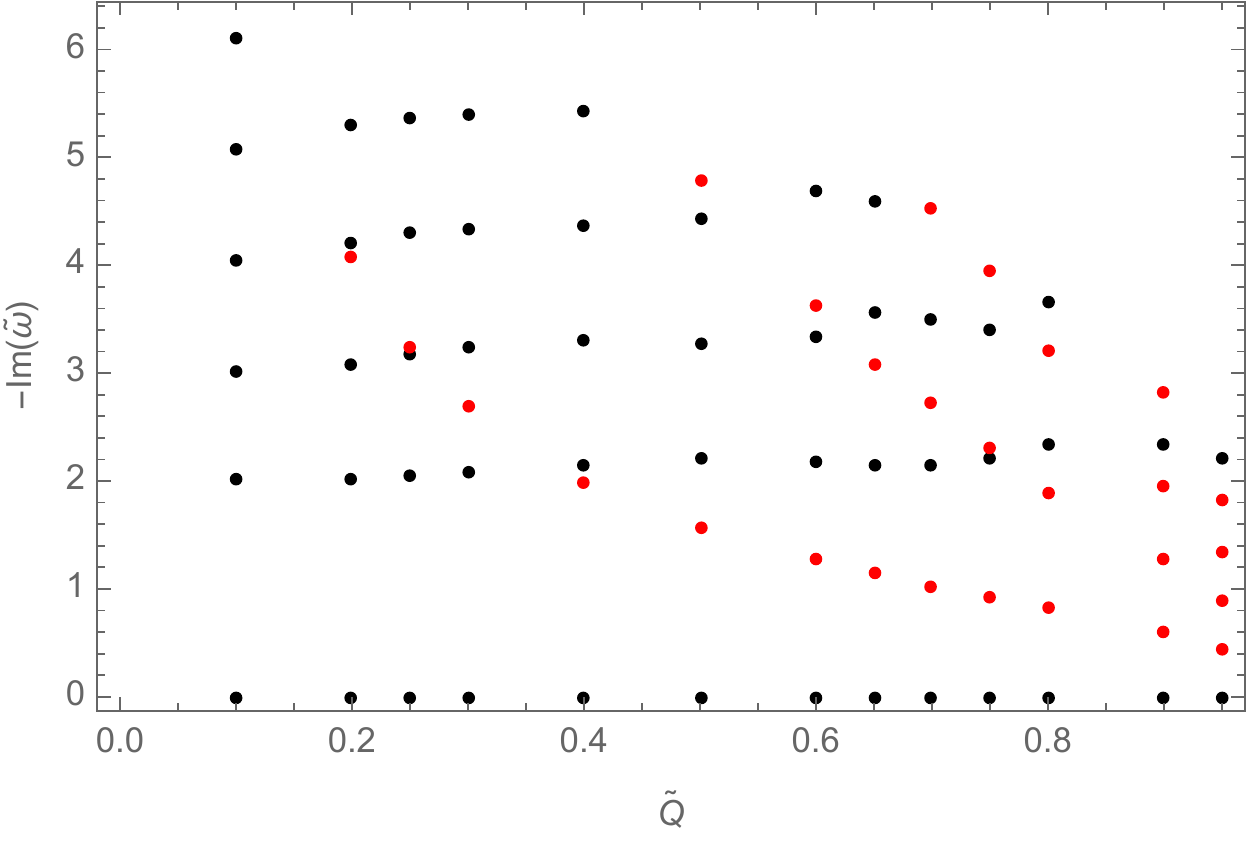}
\includegraphics[width=0.32\textwidth]{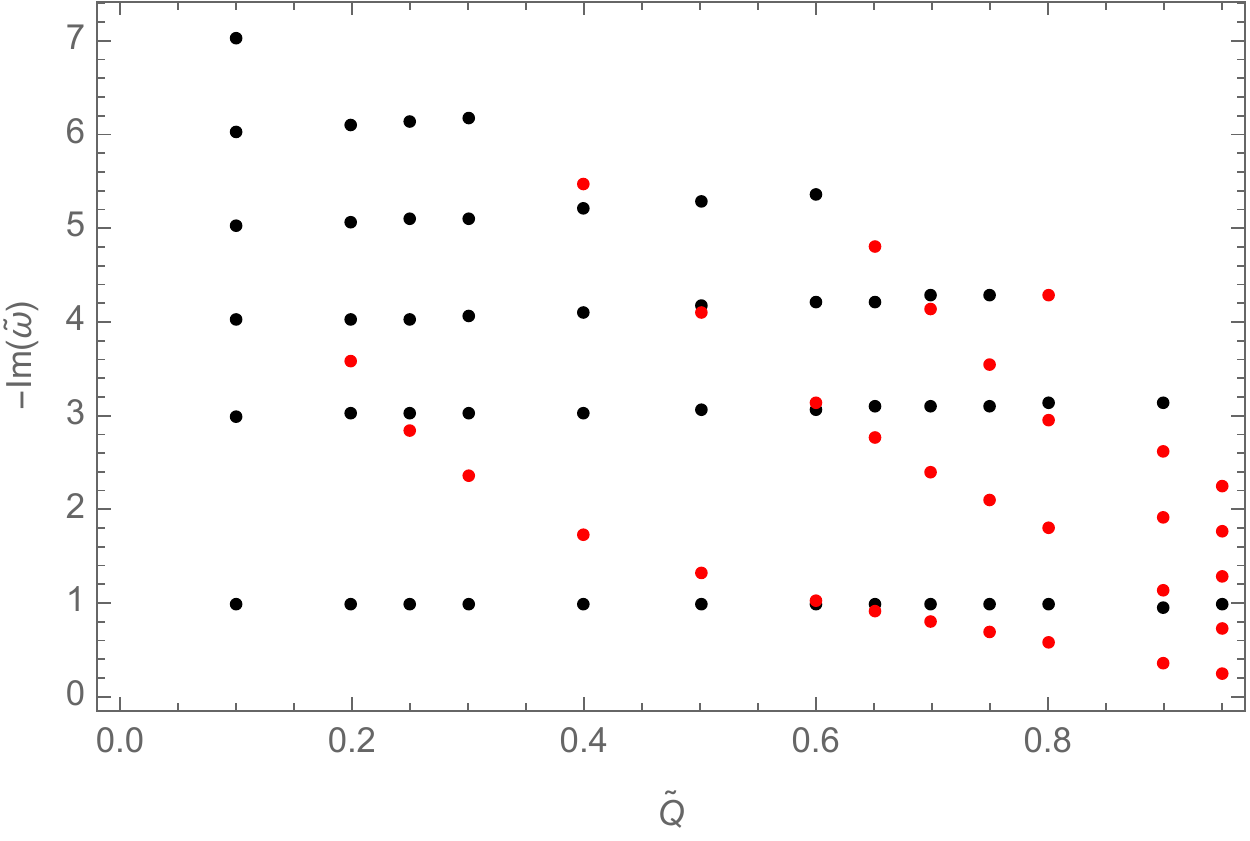}
\includegraphics[width=0.32\textwidth]{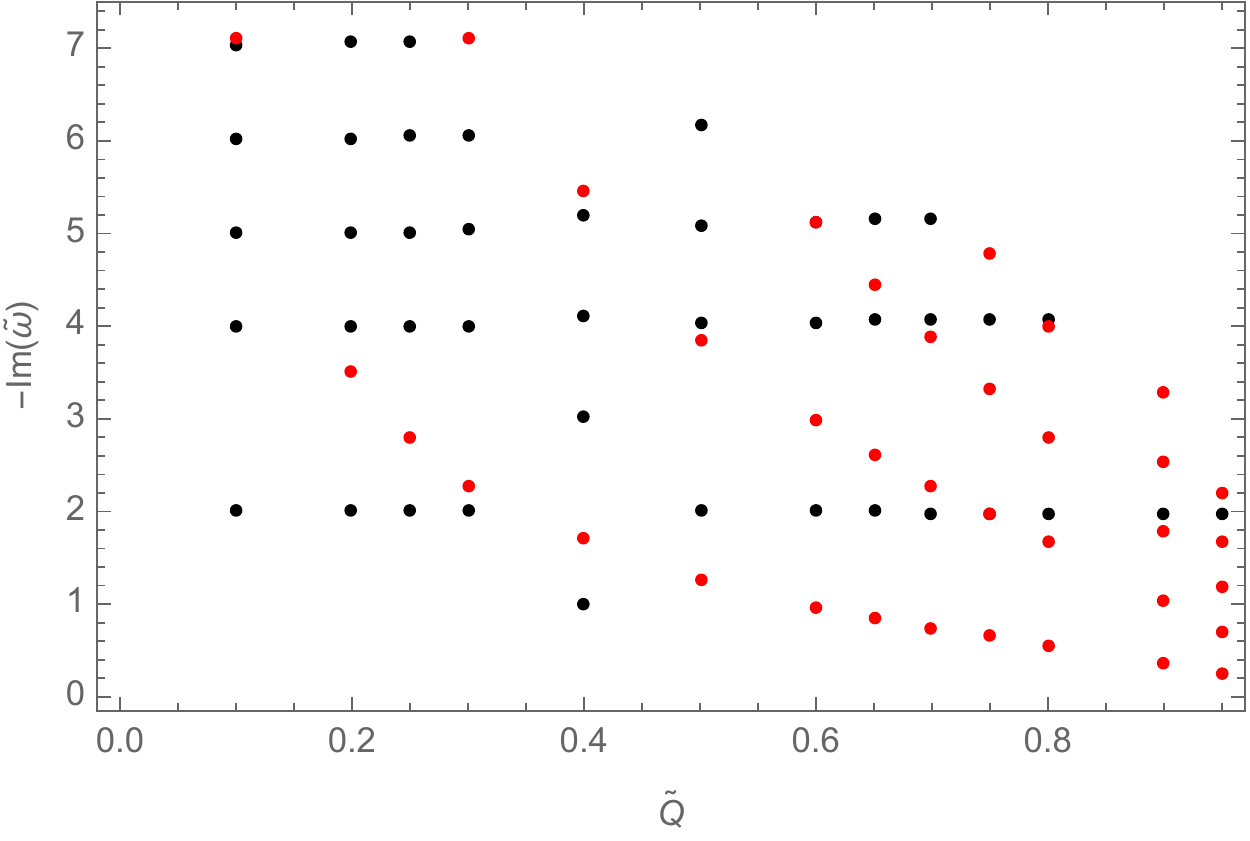}
\end{center}
\caption{$-Im(\tilde{\omega})$ as a function of $\tilde{Q}$, for massless scalar fields in the background of Weyl black holes.
Here, the QNFs are obtained via the pseudospectral Chebyshev method. Black points correspond to dS modes while that red points correspond to PS modes. Left panel for $l=0$, central panel for $l=1$, and right panel for $l=2$. Some numerical values are shown in appendix \ref{tables} Table \ref{DominanceQ}  .}
\label{DQ}
\end{figure}

As we mentioned the purely imaginary modes belong to the family of dS modes, and they continuously
approach those of pure de Sitter space in the limit that $\tilde{Q}$ vanishes. However, the dS modes also can acquire a real part if the mass of the scalar field increases enough, which is similar to what happens to the modes of pure de Sitter spacetime. The other family corresponds to complex modes, for massless and massive scalar field, with a non null real part namely PS modes.
Thus,
in order to analyze the dominance between the family modes, we plot in Fig. \ref{DQ} both families, black points correspond to dS modes and red points to PS modes. So, for massless scalar fields, we can observe that, the dS family is dominant for $l=0$. Also, the dS modes are dominant for $l=1,2$, and small values of the parameter $\tilde{Q}$. But, for higher values of $\tilde{Q}$ the PS modes are dominant for massless scalar field. 
Therefore, there is a critical value of $\tilde{Q}_c$, where for $\tilde{Q}<\tilde{Q}_c$ the  dominant family is the dS; otherwise the  dominant family is the PS, for $l>0$ and massless scalar field. 

\begin{figure}[H]
\begin{center}
\includegraphics[width=0.32\textwidth]{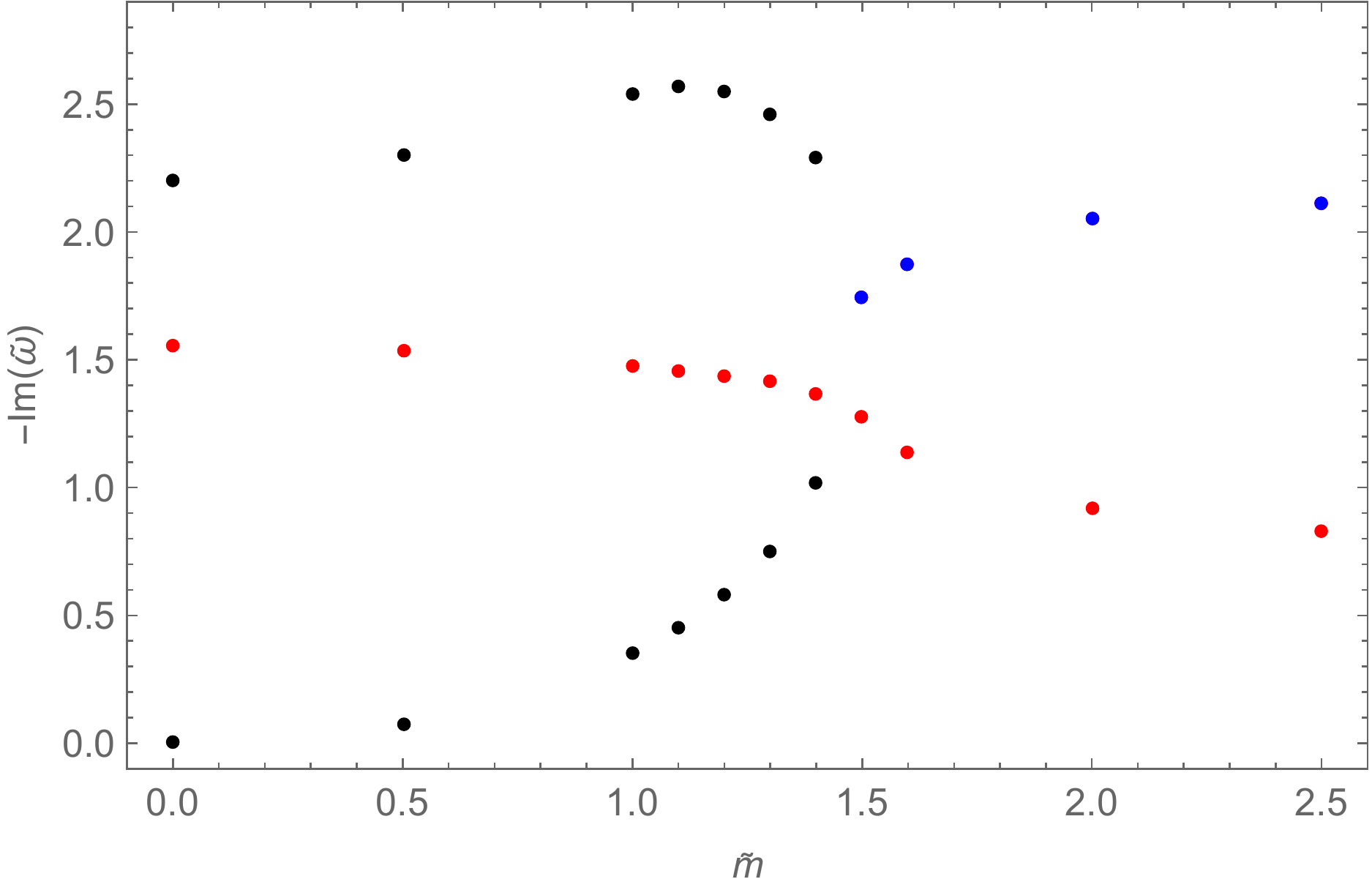}
\includegraphics[width=0.32\textwidth]{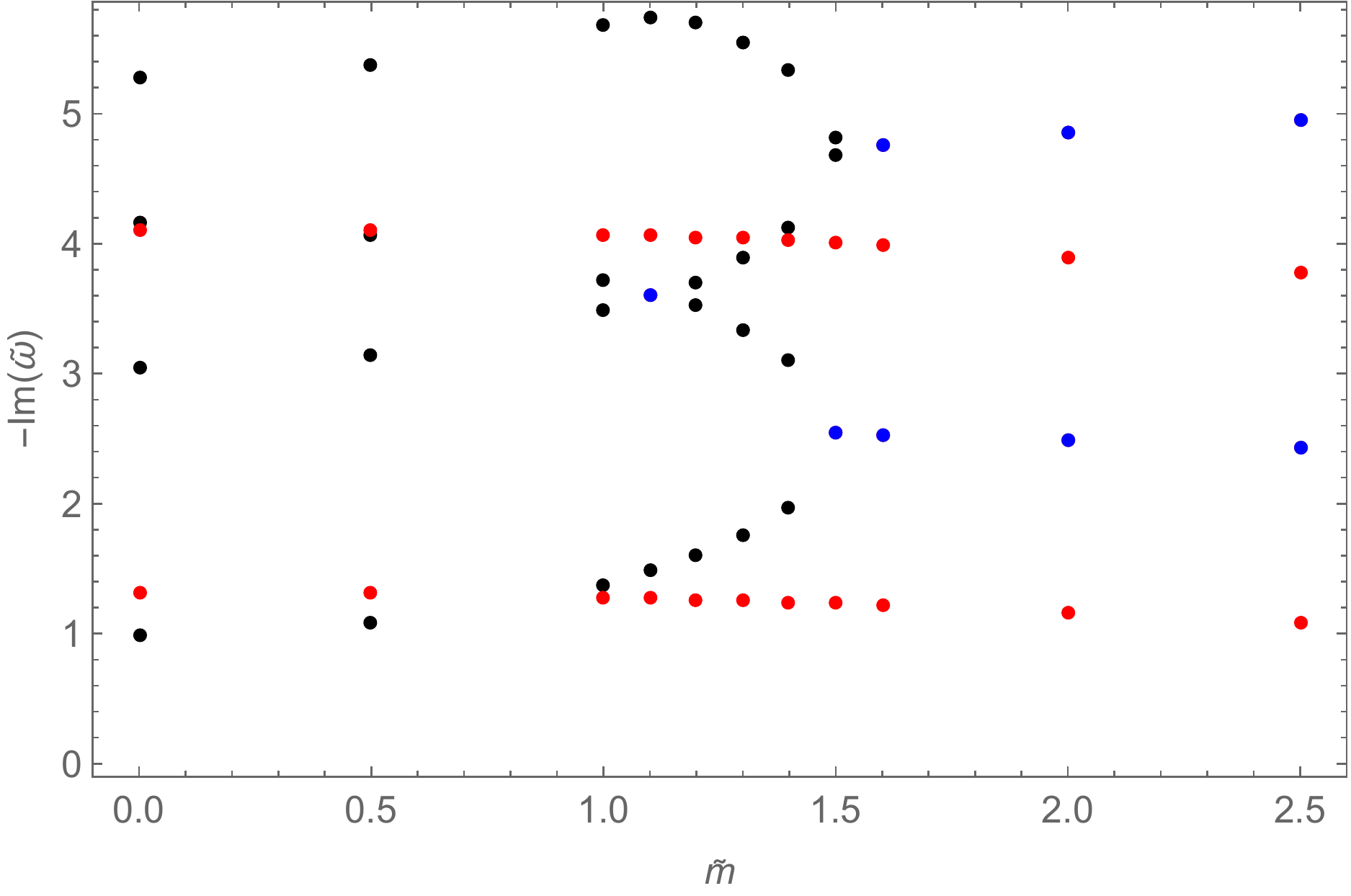}
\includegraphics[width=0.32\textwidth]{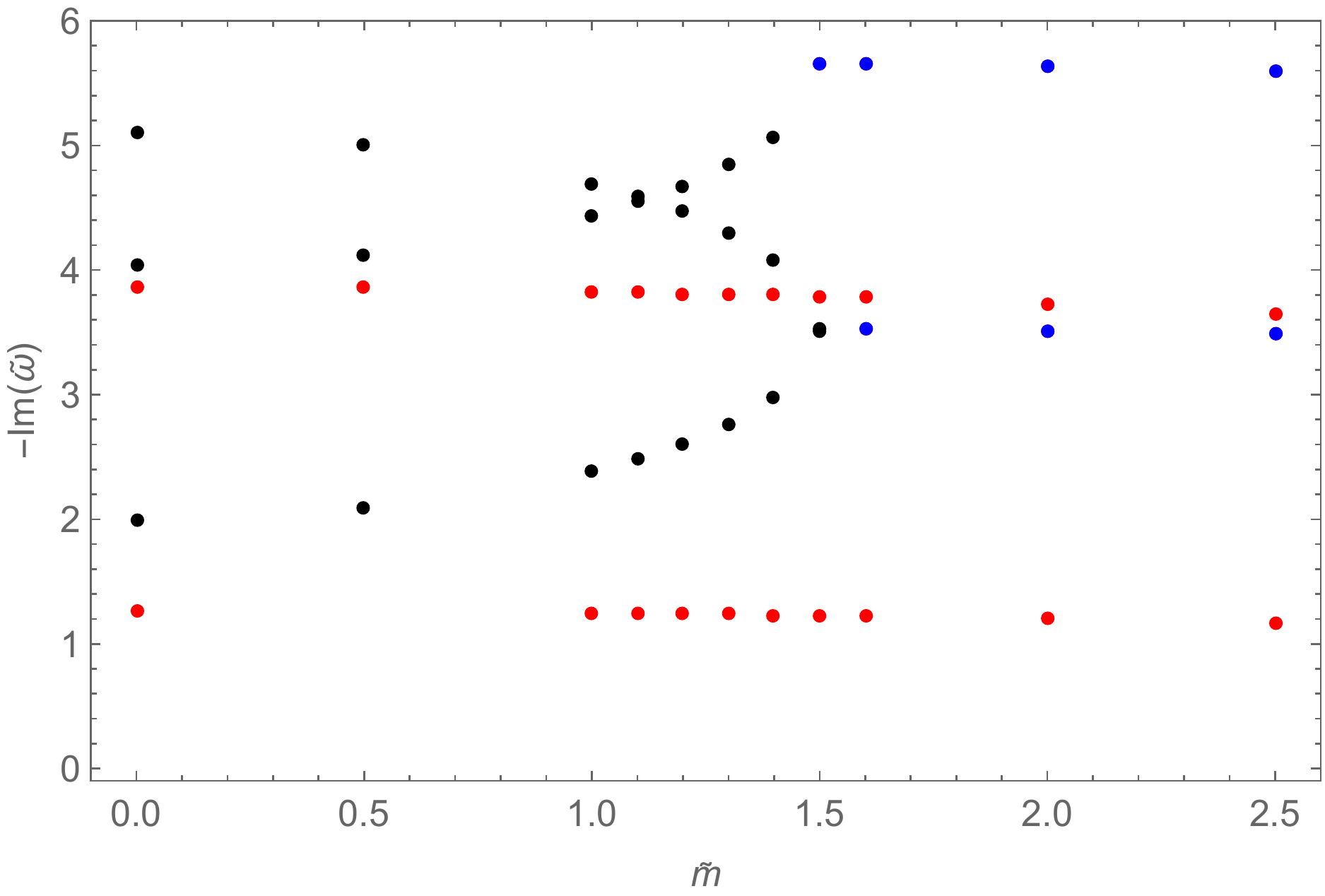}
\end{center}
\caption{$-Im(\tilde{\omega})$ as a function of $\tilde{m}$, for scalar fields in the background of Weyl black hole with $\tilde{Q}=0.5$.
Here, the QNFs are obtained via the pseudospectral Chebyshev method. Black points correspond to dS modes with a purely imaginary QNF, blue points correspond to dS modes with a complex QNF, and red points to PS modes. Left panel for $l=0$, central panel for $l=1$, and right panel for $l=2$. Some numerical values are shown in appendix \ref{tables} Table \ref{DominanceMassive1}.}
\label{DQ05m}
\end{figure}

\begin{figure}[H]
\begin{center}
\includegraphics[width=0.32\textwidth]{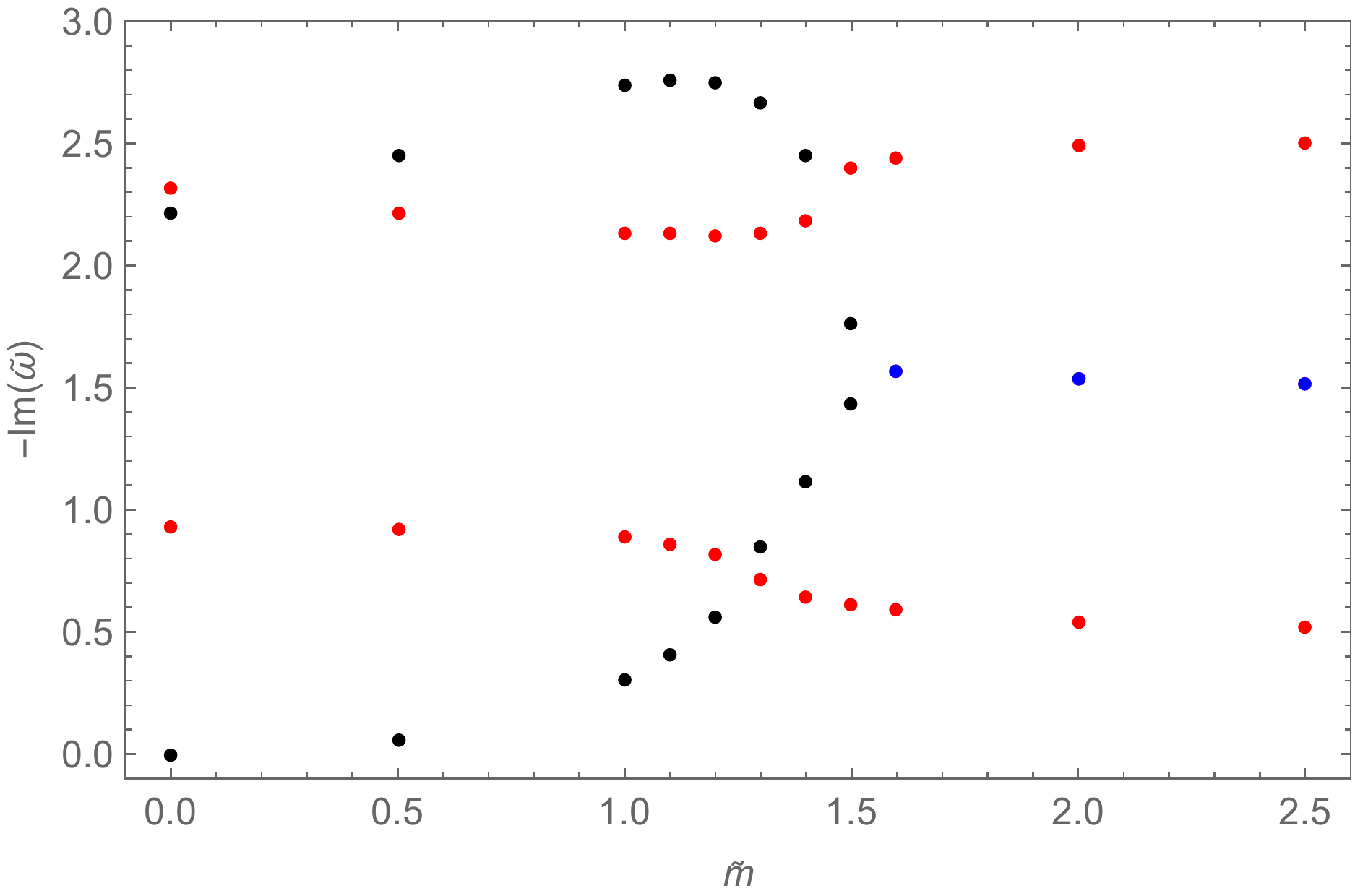}
\includegraphics[width=0.32\textwidth]{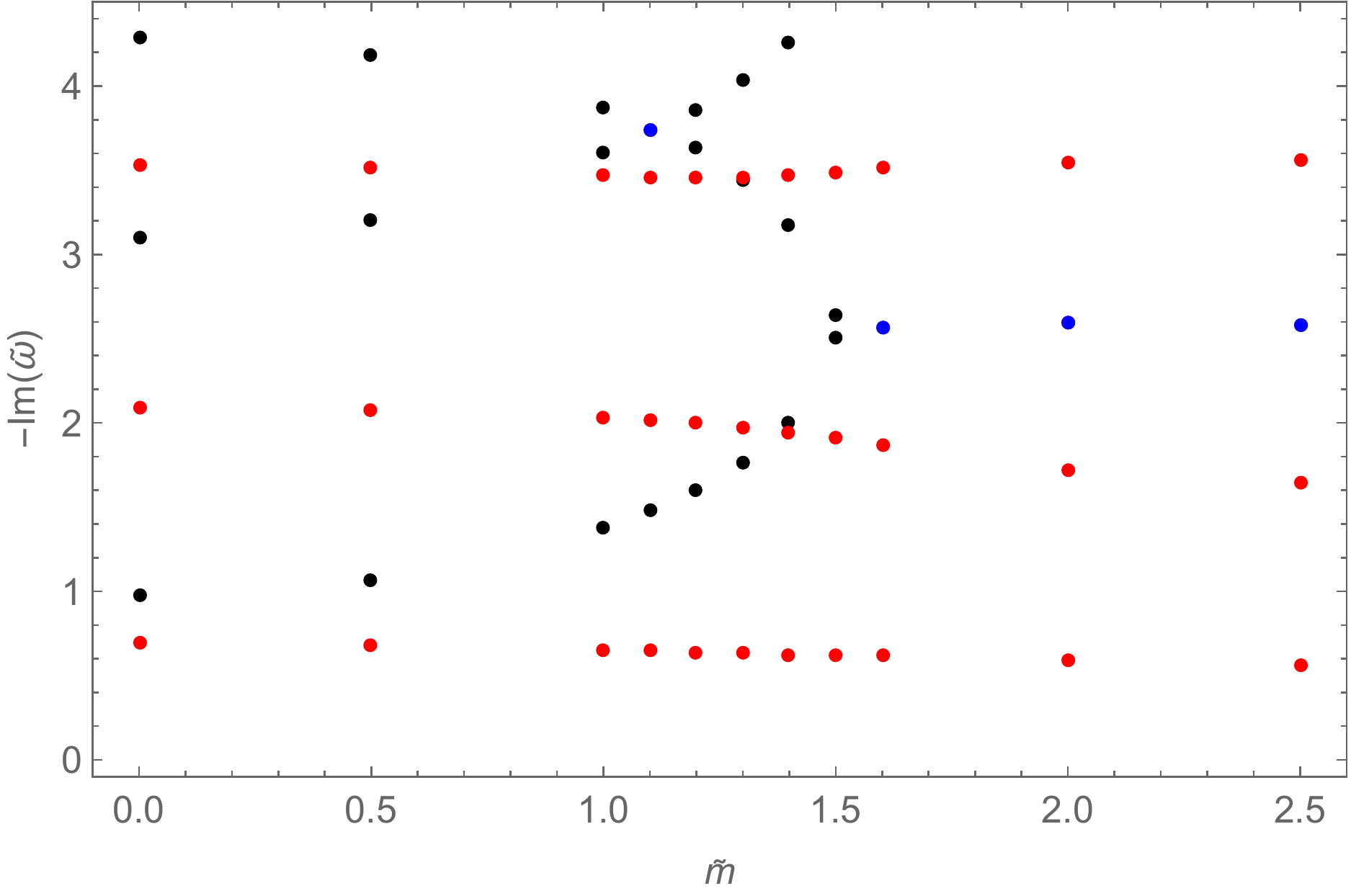}
\includegraphics[width=0.32\textwidth]{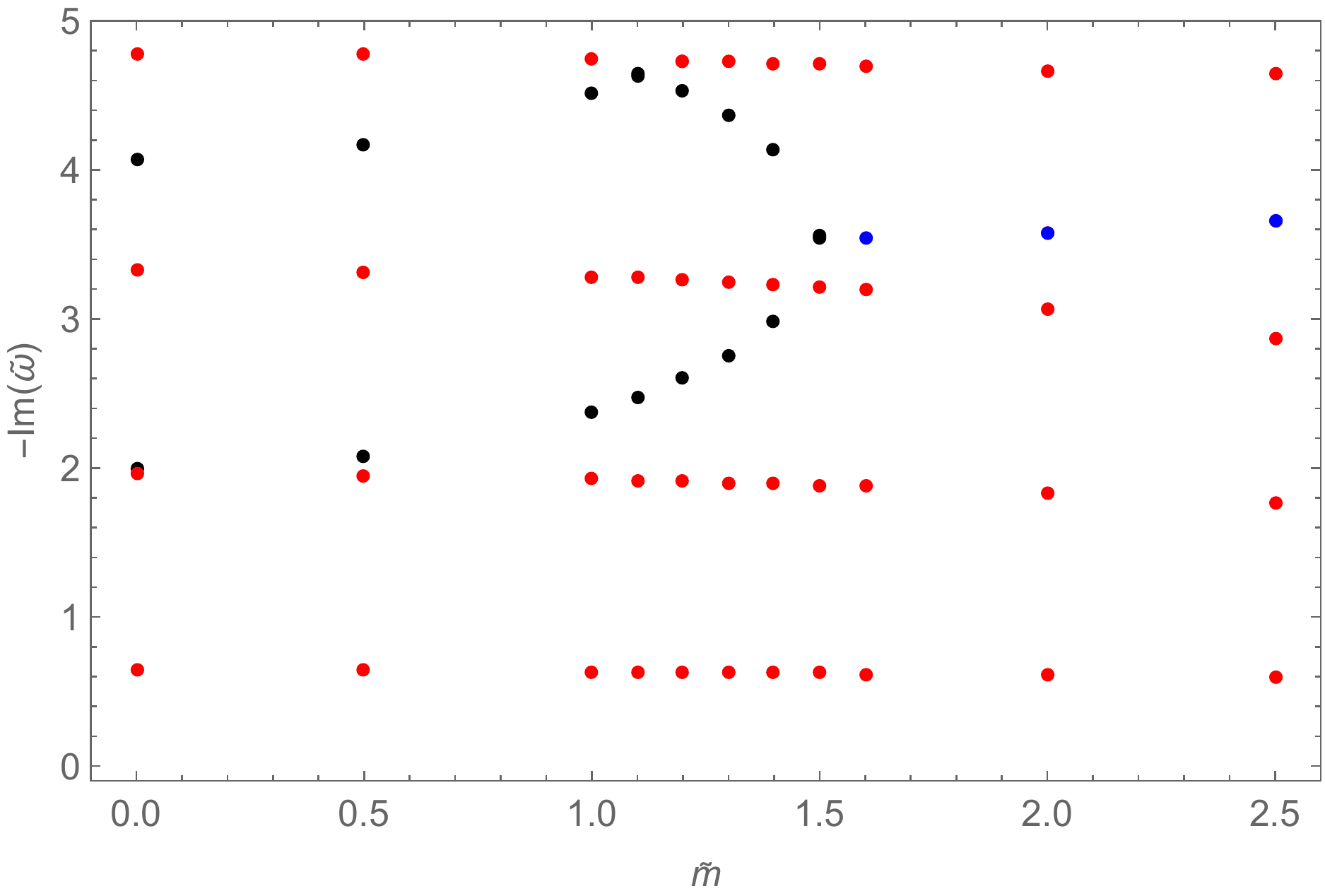}
\end{center}
\caption{$-Im(\tilde{\omega})$ as a function of $\tilde{m}$, for scalar fields in the background of Weyl black hole with $\tilde{Q}=0.75$.
Here, the QNFs are obtained via the pseudospectral Chebyshev method. Black points correspond to dS modes with a purely imaginary QNF, blue points correspond to dS modes with a complex QNF, and red points to PS modes. Left panel for $l=0$, central panel for $l=1$, and right panel for $l=2$. Some numerical values are shown in appendix \ref{tables} Table \ref{DominanceMassive2}.}
\label{DQ075m}
\end{figure}

Then, in order to analyze the effect of the scalar field mass on the dominance, we plot in Fig. \ref{DQ05m} and Fig. \ref{DQ075m} both families for $\tilde{Q}=0.5$, and $0.75$ respectively. Black points correspond to dS modes with a purely imaginary QNF, blue points correspond to dS modes with a complex QNF, and red points to PS modes. Interestingly, for $l=0$, the dominance of the dS modes depends on the scalar fields mass, for small values of $\tilde{m}$ the dS family with purely imaginary QNF is dominant. Otherwise, the PS family is dominant, see left panels. Therefore, there is a critical value of $\tilde{m}=\mu_c$, such that, for $\tilde{m}<\mu_c$, the dS modes with purely imaginary QNF are the dominant; otherwise the PS modes are dominant. Note that the dS family with complex QNF does not dominate. Also, note that the same behaviour is observed for $l=1$ and small values of $\tilde{Q}$, see central panel of Fig. \ref{DQ05m}. Remarkably, for higher values of $\tilde{Q}$, and $l>0$, the dominance of the PS family does not depend on the scalar field mass, see central and right panel of Fig.\ref{DQ075m}, and the PS is the dominant family.

Now, in order to give an approximate value of $\tilde{m}=\mu_c$, where there is an interchange in the family dominance, we consider $Im(\omega_{dS})=Im(\omega_{PS})$ as a proxy for where the interchange in the family dominance occurs, where for $\omega_{PS}$ we consider the analytical expression given by Eq. (\ref{frequency}), which yields the QNFs at third order beyond the eikonal limit, and for $\omega_{dS}$ we consider the analytical expression given by Eq. (\ref{deS}) for the pure de Sitter spacetime $\omega_{pure-dS}$ with $\Lambda_{eff} = 3/ \lambda^2$, which yields well-approximated QNFs for the dS family for high values of $l$. It is important because allows discern if the dominant family is able to suffers the anomalous behaviour of the decay rate. So, the equality of $Im(\omega_{pure-dS})$ with  $Im(\omega_{PS})$ for $n_{PS}=0$ and $n_{dS}=0$ yields
\begin{eqnarray}
\nonumber  \mu_c^2 &=& -\Bigg[ -27 + 1024 l^4 + 9 \tilde{Q}^4 \left(78-84\tilde{Q}^2+29 \tilde{Q}^4 \right) +192 l (1-\tilde{Q}^2) \left(-2(1-\tilde{Q}^2)+ 3 \sqrt{2}\tilde{Q}\sqrt{1-\tilde{Q}^2} \right) +  \\
\nonumber  && 128 l^3 \left(16+3 \sqrt{2} \tilde{Q} (1-\tilde{Q}^2)^{3/2} \right) - 64 l^2 \left(-10+ 3 \tilde{Q} \left(-5 \sqrt{2} \sqrt{1-\tilde{Q}^2} + \tilde{Q} \left(-4+ 2 \tilde{Q}^2 + 5\sqrt{2} \tilde{Q} \sqrt{1-\tilde{Q}^2} \right) \right) \right) \\
\nonumber  && - 4 \tilde{Q}^2 \Bigg( 45+ \frac{8 \sqrt{2} l(l+1)}{\tilde{Q}^2} \Bigg[ 512 l^4 +9 (1-\tilde{Q}^2)^3 (-3+7 \tilde{Q}^2) -192 l (1-\tilde{Q}^2) \left(2 (1-\tilde{Q}^2)- 3 \sqrt{2} \tilde{Q} \sqrt{1-\tilde{Q}^2} \right)   \\
\nonumber  && -128 l^3 \left(-8 - 3\sqrt{2}\tilde{Q} (1-\tilde{Q}^2)^{3/2} \right) -64 l^2 \left(-2+3 \tilde{Q} \left(-5 \sqrt{2} \sqrt{1-\tilde{Q}^2}+\tilde{Q} \left(-4+2\tilde{Q}^2 + 5\sqrt{2} \tilde{Q} \sqrt{1-\tilde{Q}^2} \right) \right) \right)  
 \\
 && \Bigg]^{1/2}   \Bigg)\Bigg] / \left( 144 \tilde{Q}^2 (1-\tilde{Q}^2)^3 \right)\,,
\end{eqnarray}
for $\mu_c \leq 3/2$, and
\begin{equation}
\mu_c^2 = -\frac{256 l^3 \tilde{Q}-3 \sqrt{2} (1-\tilde{Q}^2)^{3/2} (3+29\tilde{Q}^2)-128 l^2 \left(-5 \tilde{Q} + \sqrt{2} \sqrt{1-\tilde{Q}^2} \right) -128 l  \left(-3 \tilde{Q}+ \sqrt{2} \sqrt{1-\tilde{Q}^2} \right)}{48 \sqrt{2} \tilde{Q}^2 (1-\tilde{Q}^2)^{3/2}}
\end{equation}
for $\mu_c > 3/2$. The mass where the transition of dominance occurs $\mu_c$ depends on the parameters $\tilde{Q}$ and $l$.\\

\begin{figure}[H]
\begin{center}
\includegraphics[width=0.35\textwidth]{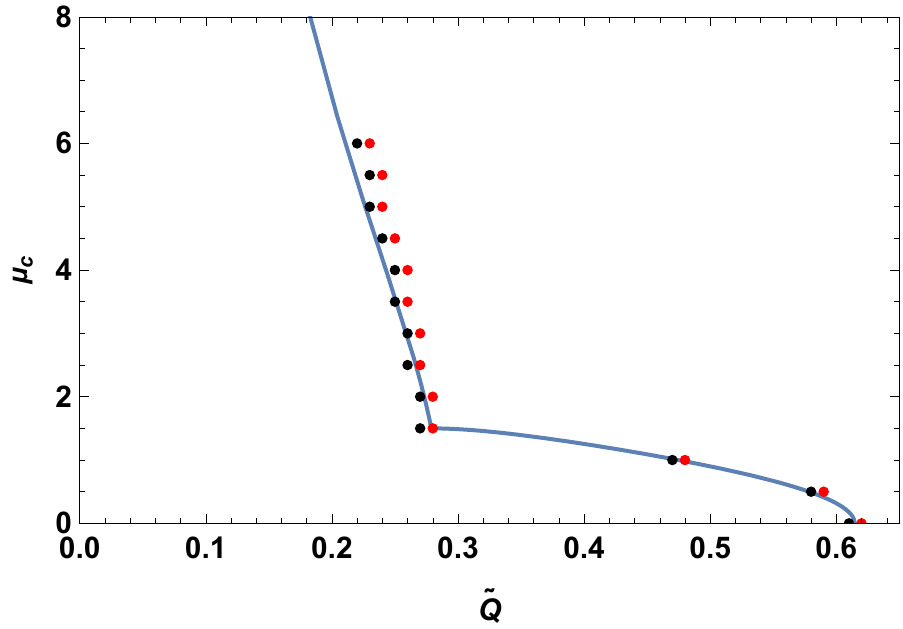}
\includegraphics[width=0.36\textwidth]{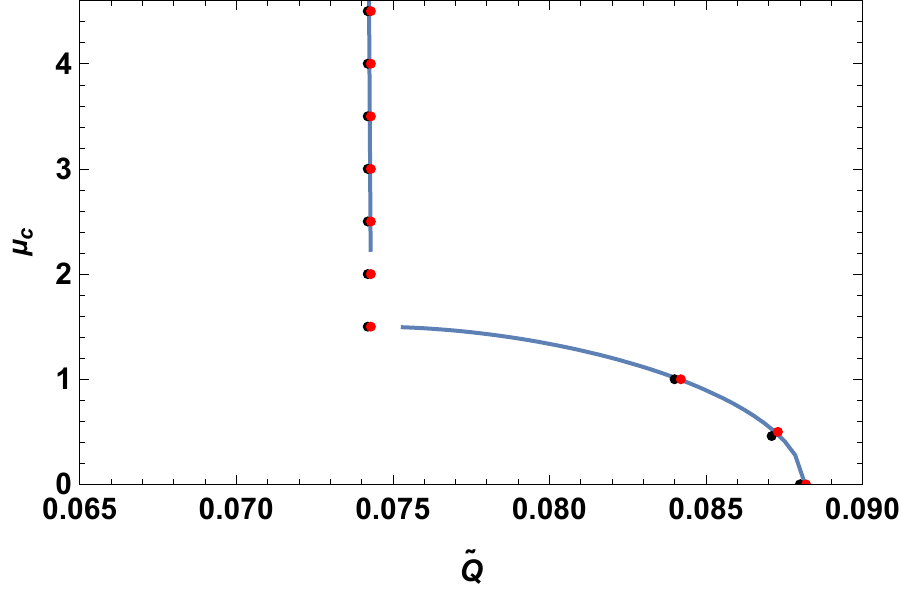}
\end{center}
\caption{The solid line corresponds to $\mu_c$ as a function of $\tilde{Q}$ for $l=1$ (left panel) and $l=8$ (right panel), and separates regions in the parameters space where a family of QNFs dominates according to the analytical approximation. In the region to the left of the line always the de Sitter modes dominate, while that in the region to the right the PS modes dominate. The numerical results using the pseudospectral Chebyshev method are represented by points. The points are close to the frontier where  the change of dominance occurs. The black points are in the side where the de Sitter modes dominate, while that the red points are in the side where the PS modes dominate. The coincidence of the line with the points for $l=1$ is more accurate for low values of $\mu_c$ while that for bigger values of $\mu_c$ the difference between the analytical and the numerical results increases. However, for $l=8$ the analytical approximation is accurate even for high values of $\mu_c$.}
\label{Dominance}
\end{figure}

Note that the imaginary part of the QNFs of the pure de Sitter depends on $m$ for $m^2 \leq 3 \Lambda_{eff}/4$ ($\tilde{m} < 3/2 $), while that in the opposite case it does not depend on $m$ and the QNFs acquire a real part. This is reflected in the behaviour of $\mu_c$, which is different for $\mu_c \leq 3/2$ and $\mu_c > 3/2$ as is shown in Fig. \ref{Dominance}. Also, in that figure we show numerical results using the pseudospectral Chebyshev method (red and blue points) where there is a change of dominance of the families. We observe that the analytical values of $\mu_c$, represented by the solid line, is more accurate for high values of $l$.

\section{Conclusions}
\label{conclusion}

In this work we studied the propagation of massive scalar fields in the Weyl black hole as a background, and we analyzed their QNFs. Mainly, we  showed that 
two families of modes are present.
One of them is a family of complex QNFs, and the other one is a family of purely imaginary modes for massless scalar fields (for massive scalar fields, the dS modes also can be complex). We showed that the purely imaginary modes belong to the family of de Sitter modes, and they continuously
approach those of pure de Sitter space in the limit that the black hole parameter $Q$ vanishes, and the complex ones corresponds to the photon sphere modes. Both families of modes show that the propagation of massless and massive scalar fields is stable in Weyl black hole backgrounds, for the cases considered. \\

For the PS modes and by using the WKB method at third order beyond the eikonal limit, we were able to estimate the value of the critical scalar field mass, and we found their dependence on $\tilde{Q}$, and on the overtone number $n_{PS}$ in the eikonal limit. Mainly, we found that the critical scalar field mass decreases when $\tilde{Q}$ increases, and it increases when the overtone number $n_{PS}$ increases. Interestingly, at the extremal limit $\tilde{Q} \rightarrow 1$ or $(r_{c}\rightarrow r_{h})$, $\tilde{m}_c \rightarrow \sqrt{2}$, and it does not depend on the overtone number $n_{PS}$. Also, when $\tilde{Q} \rightarrow 0$, $\tilde{m}_c \rightarrow \infty$. Then, we showed the anomalous decay rate of the QNMs via the WKB method at sixth order beyond the eikonal limit, where both methods i.e, the WKB method and the pseudospectral Chebyshev, show a high accuracy. Also, we showed that the frequency of oscillation increases when the scalar field mass increases, and such frequency decreases when the overtone number $n_{PS}$ or the parameter $\tilde{Q}$ increases.\\

For the dS modes we found that for massless scalar fields, and  $l$ null, the decay rate is not sensitive to the increment of $\tilde{Q}$. However, the decay rate increases when $\tilde{Q}$ increases  $n_{dS}>1$, and $l>0$. Also, when the scalar field acquires mass, we showed that  the decay rate increases when the scalar field mass increases for $l=0,1,2$, $n_{dS}=0$, and purely imaginary QNFs; however, in general this is not true for higher overtone numbers.\\

Finally, using the pseudospectral Chebyshev method we studied the dominance between the families of modes. Mainly, we showed that for massless scalar field the dS modes are dominant for $l=0$. However, for $l=1,2$, there is a critical value of $\tilde{Q}_c$, so that, if $\tilde{Q}<\tilde{Q}_c$ the dominant family is the dS; otherwise the dominant family is the PS. Interestingly, when 
the scalar field acquires mass, and for $l=0$, the dominance of the dS modes depends on the scalar fields mass, and there is a critical value of $\tilde{m}=\mu_c$, such that, for $\tilde{m}<\mu_c$, the dS modes with purely imaginary QNFs are the dominant; otherwise the PS modes are dominant. The same behaviour was  observed for $l=1$ and small values of $\tilde{Q}$. Remarkably, for higher values of $\tilde{Q}$, and $l>0$, the dominance of the PS family does not depend on the scalar field mass. Then, by considering as a proxy $Im(\omega_{dS})=Im(\omega_{PS})$ we were able to estimate the value of $\mu_c$, where there is an interchange in the family dominance for a null overtone number, this value depends on the parameters $\tilde{Q}$ and $l$.\\

It is worth to mention that despite the effective potential is negative for a range of values of $r$ for $l=0$, the propagation of massive scalar field is stable. However,  it would be interesting to extent this work to the case of charged massive scalar field, and to study the superradiance, as well as, the existence of  bound states which could to trigger an instability for $l=0$.

\acknowledgments

This work is partially supported by ANID Chile through FONDECYT Grant Nº 1220871  (P.A.G., and Y. V.). P.A.G. would like to thank the Facultad de Ciencias, Universidad de La Serena and Biblioteca Regional Gabriela Mistral for its hospitality. R.B. would like to thank the Facultad de Ingenier\'{i}a y Ciencias, Universidad Diego Portales for its hospitality.

\appendix{}

\section{Pseudospectral Chebyshev method}

In order to compute the QNFs by using the pseudospectral Chebyshev method, we solve numerically the differential equation (\ref{radial}), see for instance \cite{Boyd}. First, it is convenient to perform a change of variable in order to restrict the values of the radial coordinate to the range [0,1]. Thus, performing the change of variable $y = (r-r_h)/(r_c-r_h)$, the event horizon is located at $y=0$ and the cosmological horizon at $y=1$. Therefore, the radial equation (\ref{radial}) becomes
\begin{equation}\label{rad}
f(y) R''(y) + \left( \frac{2(r_c-r_h)f(y)}{r_h+(r_c-r_h)y} +f'(y)\right) R'(y) +\frac{(r_c-r_h)^2}{\lambda^2} \left( \frac{\tilde{\omega}^2}{f(y)} - \frac{l(l+1) \lambda^2}{(r_h+(r_c-r_h)y)^2} -\tilde{m}^2 \right)R(y) =0 \,,
\end{equation}
where $\tilde{\omega} \equiv \lambda \omega$ is the dimensionless QNF. The solution of this equation in the vicinity of the event horizon is given  by
\begin{equation}
R(y \rightarrow 0) = C_1 y^{-i (r_c-r_h) \tilde{\omega}/(\lambda f'(0))} + C_2 y^{i(r_c-r_h) \tilde{\omega}/(\lambda f'(0))}\,,
\end{equation}
where the first term represents an ingoing wave and the second represents an outgoing wave near the black hole horizon. Imposing the requirement of only ingoing waves on the horizon, we fix $C_2=0$. On the other hand, the solution to the equation in the vicinity of the cosmological horizon is given  by
\begin{equation}
R(y \rightarrow 1) = D_1 (1-y)^{-i (r_c-r_h) \tilde{\omega}/(\lambda f'(1))} + D_2 (1-y)^{i(r_c-r_h) \tilde{\omega}/(\lambda f'(1))}\,.
\end{equation}
So, imposing the requirement of only ingoing waves on the cosmological horizon requires $D_1=0$. Therefore, by considering the behaviour at the event horizon and at the cosmological horizon of the scalar field, it is possible to define the following ansatz 
$R(y) = y^{-i (r_c-r_h) \tilde{\omega}/(\lambda f'(0))} (1-y)^{i(r_c-r_h) \tilde{\omega}/(\lambda f'(1))} G(y)$, and using Eq. (\ref{rad}) an equation for the new radial function $G(y)$ is obtained, which we do not write explicitly here. Then, the solution for the function $G(y)$ is assumed to be a finite linear combination of the Chebyshev polynomials, and it is inserted into the differential equation for $G(y)$. Also, the interval [0,1] is discretized at the Chebyshev collocation points. Then, the differential equation is evaluated at each collocation point. So, a system of algebraic equations is obtained, which corresponds to a generalized eigenvalue problem, and it is solved numerically to find the QNFs.\\

\section{Some numerical values}
\label{tables}

In Table \ref{QNMC} we show some QNFs obtained via the WKB method at six order, and via the pseudospectral Chebyshev method, in order to establish the accuracity of the sixth order WKB method. Then, we show some numerical values used in the Fig. \ref{RA1} in Table  \ref{dSmodes}, for the Fig.  \ref{DQ} in Table \ref{DominanceQ}, in table \ref{DominanceMassive1} for the Fig. \ref{DQ05m}, and in table \ref{DominanceMassive2} for the Fig. \ref{DQ075m}.

\begin {table}[h]
\caption {Photon sphere modes $\tilde{\omega}_{PS}$ for scalar fields with $l = 1,5,10,20,40,60$, $n_{PS}=0,1,2$, and $\tilde{m}=0,1,2$, in the background of Weyl black hole with $\tilde{Q}=0.5$. Here, the QNFs were obtained via the pseudospectral Chebyshev method with a number of Chebyshev polynomials in the range 95-100, and nine decimals places of accuracy.}
\label {QNMC}\centering
\scalebox{0.65} {
\begin {tabular} { | c | c | c | c | c |c |}
\hline
\multicolumn{6}{ |c| }{$n_{PS}=0$} \\ \hline
$l$ &  $\tilde{m}$ & \text{ Chevyshev method}  & $ WKB $ & $ \epsilon_{Re(\tilde{\omega})} (\%) $ & $ \epsilon_{Im(\tilde{\omega})} (\%) $\\\hline
$1$ & $0$ &
$\pm 2.515818188 - 1.321721786 i$ & $\pm 2.508946469 - 1.342240052 i$ & $0.273$
& $1.552$ \\
$1$ & $1.0$ &
$\pm 2.573402279 - 1.283591710 i$ & $\pm 2.565530630-1.305412427 i$ & $0.306$
&  $1.700$\\
$1$ & $2.0$ &
$\pm 2.758096393 - 1.168931505 i$ & $ \pm 2.744354026-1.195028854 i$ & $0.498$
& $2.233$ \\
\hline

$5$ & $0$ &
$\pm 9.497142740 - 1.232055590 i$ & $ \pm  9.497143016-1.232055316 i$ & $2.906\cdot10^{-6}$
& $2.224\cdot10^{-5}$ \\
$5$ & $1.0$ &
$\pm 9.516613878 - 1.229191786 i$ & $\pm 9.516614159-1.229191486 i$ & $2.953\cdot 10^{-6}$
& $2.442\cdot 10^{-5}$ \\
$5$ & $2.0$ &
$\pm 9.574990966 - 1.220662282 i$ & $\pm 9.574991261-1.220661899 i$ & $3.081\cdot 10{-6}$
& $3.138\cdot 10^{-5}$ \\
\hline

$10$ & $0$ &
$\pm 18.171122987 - 1.226747673 i$ & $\pm 18.171122990-1.226747629 i$ & $1.651\cdot 10{-8}$
& $3.587\cdot 10^{-6}$ \\
$10$ & $1.0$ &
$\pm 18.181403278 - 1.225965241 i$ & $\pm 18.181403280-1.225965198 i$ & $1.100\cdot 10^{-8}$
& $3.507\cdot 10^{-6}$ \\
$10$ & $2.0$ &
$\pm 18.212236358 - 1.223622843 i$ & $\pm 18.212236360-1.223622800 i$ & $1.098\cdot 10^{-8}$
& $3.514\cdot 10^{-6}$ \\
\hline

$20$ & $0$ &
$\pm 35.499127829 - 1.225270050 i$ & $\pm 35.499127829-1.225270050 i$ & $0$
& $0$ \\
$20$ & $1.0$ &
$\pm 35.504404534 - 1.225065043 i$ & $\pm 35.504404534-1.225065042 i$ & $0$
& $8.163\cdot 10^{-8}$ \\
$20$ & $2.0$ &
$\pm 35.520233506 - 1.224450363 i$ & $\pm 35.520233506-1.224450363 i$ & $0$
& $0$ \\
\hline

$40$ & $0$ &
$\pm 70.144049267 - 1.224879411 i$ & $\pm 70.144049267-1.224879411 i$ & $0$
& $0$ \\
$40$ & $1.0$ &
$\pm 70.146721675 - 1.224826904 i$ & $\pm 70.146721675-1.224826904 i$ & $0$
& $0$ \\
$40$ & $2.0$ &
$\pm 70.154738748 - 1.224669405 i$ & $\pm 70.154738748-1.224669405 i$ & $0$
& $0$ \\
\hline

$60$ & $0$ &
$\pm 104.786390176 - 1.224805161 i$ & $\pm 104.786390176-1.224805161 i$ & $0$
& $0$ \\
$60$ & $1.0$ &
$\pm 104.788179332 - 1.224781632 i$ & $\pm 104.788179332-1.224781632 i$ & $0$
& $0$ \\
$60$ & $2.0$ &
$\pm 104.793546754 - 1.224711052 i$ & $\pm 104.793546754-1.224711052 i$ & $0$
& $0$ \\
\hline

\multicolumn{6}{ |c| }{$n_{PS}=1$} \\ \hline
$l$ &  $\tilde{m}$ & \text{ Chevyshev method}  & $ WKB $ & $ \epsilon_{Re(\tilde{\omega})} (\%) $ & $ \epsilon_{Im(\tilde{\omega})} (\%) $\\\hline
$1$ & $0$ &
$\pm 2.088713589 - 4.105947399 i$ & $\pm 2.098260792-4.126829472 i$ & $0.457$
& $0.509$ \\
$1$ & $1.0$ &
$\pm 2.082836938 - 4.066865363 i$ & $\pm 2.091773990-4.088288772 i$ & $0.429$
& $0.527$ \\
$1$ & $2.0$ &
$\pm 1.368644128 - 2.491823896 i$ & $\pm  1.983426658-4.026998394 i$ & $44.919$
& $61.609$ \\
\hline

$5$ & $0$ &
$\pm 9.324127333 - 3.714511812 i$ & $\pm 9.324120826-3.714500174 i$ & $6.979\cdot 10^{-5}$
& $3.133\cdot 10^{-4}$ \\
$5$ & $1.0$ &
$\pm 9.341711948 - 3.706086794 i$ & $\pm 9.341705019-3.706073745 i$ & $7.417\cdot10^{-5}$
& $3.521\cdot 10{-4}$ \\
$5$ & $2.0$ &
$\pm 9.394571774 - 3.680896648 i$ & $\pm 9.394563874-3.680879235 i$ & $8.409\cdot 10^{-5}$
& $4.731\cdot 10^{-4}$ \\
\hline

$10$ & $0$ &
$\pm 18.078886488 - 3.685245260 i$ & $\pm 18.078886386-3.685244664 i$ & $5.642\cdot 10^{-7}$
& $1.617\cdot10^{-5}$ \\
$10$ & $1.0$ &
$\pm 18.088910524 - 3.682906299 i$ & $\pm 18.088910421-3.682905695 i$ & $5.694\cdot10^{-7}$
& $1.640\cdot10^{-5}$ \\
$10$ & $2.0$ &
$\pm 18.118981654 - 3.675901945 i$ & $\pm 18.118981549-3.675901319 i$ & $5.795\cdot10^{-7}$
& $1.703\cdot 10^{-5}$ \\
\hline

$20$ & $0$ &
$\pm 35.451674563 - 3.677118196 i$ & $\pm 35.451674562-3.677118183 i$ & $2.821\cdot10^{-9}$
& $3.535\cdot10^{-7}$ \\
$20$ & $1.0$ &
$\pm 35.456917464 - 3.676503668 i$ & $\pm 35.456917463-3.676503655 i$ & $2.820\cdot10^{-9}$
& $3.536\cdot10^{-7}$ \\
$20$ & $2.0$ &
$\pm 35.472645275 - 3.674661067 i$ & $\pm 35.472645274-3.674661055 i$ & $2.819\cdot10^{-9}$
& $3.266\cdot10^{-7}$ \\
\hline

$40$ & $0$ &
$\pm 70.120002353 - 3.674973044 i$ & $\pm 70.120002353-3.674973043 i$ & $0$
& $2.721\cdot10^{-8}$ \\
$40$ & $1.0$ &
$\pm 70.122670399 - 3.674815553 i$ & $\pm 70.122670399-3.674815553 i$ & $0$
& $0$ \\
$40$ & $2.0$ &
$\pm 70.130674396 - 3.674343147 i$ & $\pm 70.130674396-3.674343147 i$ & $0$
& $0$ \\
\hline

$60$ & $0$ &
$\pm 104.770289199 - 3.674565490 i$ & $\pm 104.770289198-3.674565490 i$ & $9.545\cdot10^{-10}$
& $0$ \\
$60$ & $1.0$ &
$\pm 104.772077047 - 3.674494911 i$ & $\pm 104.772077047-3.674494911 i$ & $0$
& $0$ \\
$60$ & $2.0$ &
$\pm 104.777440549 - 3.674283189 i$ & $\pm 104.777440549-3.674283189 i$ & $0$
& $0$ \\
\hline

\multicolumn{6}{ |c| }{$n_{PS}=2$} \\ \hline
$l$ &  $\tilde{m}$ & \text{ Chevyshev method}  & $ WKB $ & $ \epsilon_{Re(\tilde{\omega})} (\%) $ & $ \epsilon_{Im(\tilde{\omega})} (\%) $\\\hline
$1$ & $0$ &
$\pm 1.561695815 - 7.368587767 i $ & $\pm 1.335061389-7.572242514 i$ & $14.512$
& $2.764$ \\
$1$ & $1.0$ &
$ \pm 1.561163041 - 7.343999390 i$ & $\pm 1.384060455-7.583399611 i$ & $11.344$
& $3.260$ \\
$1$ & $2.0$ &
$\pm 2.124422533 - 3.896668872 i$ & $\pm 1.268113517-8.493321065 i$ & $40.308$
& $117.964$ \\
\hline

$5$ & $0$ &
$\pm 8.981055931 - 6.255145178 i$ & $\pm 8.980686808-6.254735733 i$ & $4.110\cdot10^{-3}$
& $6.546\cdot10^{-3}$ \\
$5$ & $1.0$ &
$\pm 8.994609175 - 6.242027275 i$ & $\pm 8.994187221-6.241575927 i$ & $4.691\cdot10^{-3}$
& $7.231\cdot10^{-3}$ \\
$5$ & $2.0$ &
$\pm 9.035451051 - 6.202582881 i$ & $\pm 9.034858016-6.201985662 i$ & $6.563\cdot10^{-3}$
& $9.629\cdot10^{-3}$ \\
\hline

$10$ & $0$ &
$\pm 17.894551052 - 6.159044973 i$ & $\pm 17.894545446-6.159031007 i$ & $3.133\cdot10^{-5}$
& $2.268\cot10^{-4}$ \\
$10$ & $1.0$ &
$\pm 17.904043780 - 6.155183051 i$ & $\pm 17.904037994-6.155168698 i$ & $3.232\cdot10^{-5}$
& $2.332\cdot10^{-4}$ \\
$10$ & $2.0$ &
$\pm 17.932532376 - 6.143611329 i$ & $\pm 17.932526045-6.143595789 i$ & $3.530\cdot10^{-5}$
& $2.529\cdot10^{-4}$ \\
\hline

$20$ & $0$ &
$\pm 35.356774191 - 6.132911934 i$ & $\pm 35.356774134-6.132911648 i$ & $1.612\cdot10^{-7}$
& $4.663\cdot10^{-6}$ \\
$20$ & $1.0$ &
$\pm 35.361948747 - 6.131889520 i$ & $\pm 35.361948690-6.131889231 i$ & $1.612\cdot10^{-7}$
& $4.713\cdot10^{-6}$ \\
$20$ & $2.0$ &
$\pm 35.377472005 - 6.128823787 i$ & $\pm 35.377471947-6.128823492i$ & $1.639\cdot10^{-7}$
& $4.813\cdot10^{-6}$ \\
\hline

$40$ & $0$ &
$\pm 70.071908890 - 6.126072533 i$ & $\pm 70.071908890-6.126072528 i$ & $0$
& $8.162\cdot10^{-8}$ \\
$40$ & $1.0$ &
$\pm 70.074568188 - 6.125810155 i$ & $\pm 70.074568188-6.125810150 i$ & $0$
& $8.162\cdot10^{-8}$ \\
$40$ & $2.0$ &
$\pm 70.082545956 - 6.125023128 i$ & $\pm 70.082545955-6.125023123 i$ & $1.427\cdot10^{-9}$
& $8.163\cdot10^{-8}$ \\
\hline

$60$ & $0$ &
$\pm 104.738087328 - 6.124776132 i$ & $\pm 104.738087328-6.124776131 i$ & $0$
& $1.633\cdot10^{-8}$ \\
$60$ & $1.0$ &
$\pm 104.739872559 - 6.124658521 i$ & $\pm 104.739872559-6.124658521 i$ & $0$
& $0$ \\
$60$ & $2.0$ &
$\pm 104.745228210 - 6.124305713 i$ & $\pm 104.745228210-6.124305712 i$ & $0$
& $1.633\cdot10^{-8}$ \\
\hline

\end {tabular}
}
\end {table}

\begin{table}[H]
\caption {de Sitter modes $\tilde{\omega}_{dS}$ for massless scalar fields in the background of Weyl black holes, with  $\tilde{Q} = 0.25, 0.50, 0.60, 0.65, 0.75$, and $0.95$. Here, the QNFs are obtained via the pseudospectral Chebyshev method using a number of Chebyshev polynomials in the range 95-100, 
%and the values with $"..."$, means that the QNFs converges for iterations in the range 165-170,
with eight decimals places of accuracy for the QNF.}
\label{dSmodes}\centering
\scalebox{0.8} {
\begin {tabular} { | c | c | c | c | c | c | c |}
\hline
${}$ & $\tilde{Q} = 0.25$ & $ \tilde{Q} = 0.50$  & $ \tilde{Q} = 0.60$  &  $ \tilde{Q} = 0.65$  & $\tilde{Q} = 0.75$ & $\tilde{Q} = 0.95$ \\\hline
$\tilde{\omega}_{dS} (l=0)$ &
$0$ &
$0$ &
$0$  &
$0$  &
$0$  &
$0$ 
 \\\hline
$\tilde{\omega}_{dS} (l=0)$ &
$-2.04671144 i$ &
$-2.20642433 i$ &
$-2.18901577 i$  &
$-2.16153418 i$  &
$-2.21473623 i$  &
$-2.21035228 i$   \\\hline
$\tilde{\omega}_{dS} (l=0)$ &
$-3.16004383 i$ &
$-3.28026122 i$ &
$-3.32448547 i$  &
$-3.55081948 i$  &
$-3.41123404 i$  &
$-3.68220838 i$     \\\hline
$\tilde{\omega}_{dS} (l=0)$ &
$-4.28933168 i$  &
$-4.43842018 i$ &
$-4.67259840 i$  &
$-4.59395116 i$  &
$-4.78100031 i$  &
$-4.89259370 i$    
%\\\hline
%$\tilde{\omega} (\ell=0)$ &
%$-5.37442907 i$ &
%$...$ &
%$...$  &
%$...$  &
%$...$  &
%$...$     \\\hline
%$\tilde{\omega} (\ell=0)$ &
%$...$ &
%$...$  &
%$...$  &
%$...$  &
%$...$  &
%$...$  
\\\hline\hline

$\tilde{\omega}_{dS} (l=1)$ &
$-0.99744235 i$ &
$-0.99061127 i$ &
$-0.98752196 i$  &
$-0.98614568 i$  &
$-0.98353510 i$  &
$-0.97323017 i$  \\\hline

$\tilde{\omega}_{dS} (l=1)$ &
$-3.01566990 i$  &
$-3.05278804 i$  &
$-3.07819966 i$  &
$-3.08861770 i$  &
$-3.10011501 i$  &
$-3.14073605 i$     \\\hline

$\tilde{\omega}_{dS} (l=1)$ &
$-4.04506503 i$ &
$-4.15918692 i$ &
$-4.19874786 i$  &
$-4.21838805 i$  &
$-4.28198250 i$  &
$-4.41954574 i$   \\\hline

$\tilde{\omega}_{dS} (l=1)$ &
$-5.08496946 i$ &
$-5.27886988 i$  &
$-5.36433739 i$  &
$-5.42848311 i$  &
$-5.52931897 i$  &
$-5.69497129 i$     
%\\\hline

%$\tilde{\omega} (\ell=1)$ &
%$-6.13474436 i$ &
%$...$ &
%$...$  &
%$...$  &
%$...$  &
%$...$     \\\hline

%$\tilde{\omega} (\ell=1)$ &
%$...$ &
%$...$  &
%$...$  &
%$...$  &
%$...$  
\\\hline\hline

$\tilde{\omega}_{dS} (l=2)$ &
$-1.99896780 i$ &
$-1.99595932 i$  &
$-1.99424597 i$  &
$-1.99329952 i$  &
$-1.99126148 i$  &
$-1.98621495 i$     \\\hline

$\tilde{\omega}_{dS} (l=2)$ &
$-4.00897172 i$  &
$-4.03347941 i$ &
$-4.04665462 i$  &
$-4.05405720 i$  &
$-4.06889645 i$  &
$-4.10284182 i$     \\\hline

$\tilde{\omega}_{dS} (l=2)$ &
$-5.02743297 i$ &
$-5.09907861 i$  &
$-5.13697095 i$  &
$-5.15597703 i$  &
$-5.19877493 i$  &
$-5.29010803 i$    \\\hline

$\tilde{\omega}_{dS} (l=2)$ &
$-6.05413248 i$ &
$-6.18973980 i$ &
$-6.25575481 i$  &
$-6.29309604 i$  &
$-6.36824958 i$  &
$-6.52688930 i$    

\\\hline
\end {tabular}
}
\end{table}\leavevmode\newline

\begin{table}[H]
\caption {Quasinormal frequencies $\tilde{\omega}$ for massless scalar fields in the background of Weyl black holes, with $\tilde{Q} = 0.25, 0.50, 0.75$. Here, the QNFs are obtained via the pseudospectral Chebyshev method using a number of Chebyshev polynomials in the range 95-100, 
%and the values with $"..."$, means that the QNFs converges for iterations in the range 165-170,
with eight decimals places of accuracy for the QNF.}
\label{DominanceQ}\centering
\scalebox{0.65} {

\begin {tabular} { | c | c | c | c | c | c | c |}
\hline
${}$ & $\tilde{Q} = 0.25$ & $ \tilde{Q} = 0.50$  & $ \tilde{Q} = 0.60$  &  $ \tilde{Q} = 0.65$  & $\tilde{Q} = 0.75$ & $\tilde{Q} = 0.95$ \\\hline
$\tilde{\omega} (l=0)$ &
$0$ &
$0$ &
$0$  &
$0$  &
$0$  &
$0$ 
 \\\hline
$\tilde{\omega} (l=0)$ &
$-2.04671144 i$ &
$\pm 0.96483271 - 1.55503332 i$ &
$\pm 0.74231895 - 1.26271835 i$  &
$\pm 0.64478757 - 1.14423829 i$  &
$\pm 0.46343497 - 0.93126623 i$   &
$\pm 0.10726001 - 0.42652388 i$ \\\hline
$\tilde{\omega} (l=0)$ &
$-3.16004383 i$ &
$-2.20642433 i$ &
$-2.18901577 i$  &
$-2.16153418 i$  &
$-2.21473623 i$  &
$\pm 0.16969121 - 0.88043287 i$   \\\hline
$\tilde{\omega} (l=0)$ &
$\pm 2.12783506 - 3.24636263 i$ &
$-3.28026122 i$ &
$-3.32448547 i$  &
$\pm 0.47884563 - 3.09352762 i$  &
$\pm 0.33711158 - 2.31236354 i$  &
$\pm 0.18822951 - 1.34224560 i$   \\\hline
$\tilde{\omega} (l=0)$ &
$-4.28933168 i$ &
$-4.43842018 i$ &
$\pm 0.51212397 - 3.62177318 i$  &
$-3.55081948 i$  &
$-3.41123404 i$  &
$\pm 0.16402991 - 1.80805408 i$   \\\hline
$\tilde{\omega} (l=0)$ &
$-5.37442907 i$ &
$\pm 0.65424586 - 4.76952419 i$ &
$-4.67259840 i$  &
$-4.59395116 i$  &
$\pm 0.32713306 - 3.94683557 i$ &
$-2.21035228 i$    \\\hline\hline

$\tilde{\omega} (l=1)$ &
$-0.99744235 i$ &
$-0.99061127 i$ &
$-0.98752196 i$  &
$\pm 1.64157770 - 0.91377744 i$  &
$\pm 1.19557672 - 0.69368640 i$ &
$\pm 0.40857980 - 0.24093624 i$    \\\hline

$\tilde{\omega} (l=1)$ &
$\pm 5.77602977 - 2.85380824 i$ &
$\pm 2.51581819 - 1.32172179 i$ &
$\pm 1.89754750 - 1.03464690 i$  &
$-0.98614568 i$  &
$-0.98353510 i$  &
$\pm 0.37430664 - 0.73992536 i$   \\\hline

$\tilde{\omega} (l=1)$ &
$-3.01566990 i$ &
$-3.05278804 i$ &
$-3.07819966 i$  &
$\pm 1.47078540 - 2.75715839 i$  &
$\pm 1.13539826 - 2.08983835 i$  &
$-0.97323017 i$   \\\hline

$\tilde{\omega} (l=1)$ &
$-4.04506503 i$ &
$\pm 2.08871359 - 4.10594740 i$ &
$\pm 1.65290106 - 3.14670689 i$  &
$-3.08861770 i$  &
$-3.10011501 i$  &
$\pm 0.37733607 - 1.29048473 i$    \\\hline

$\tilde{\omega} (l=1)$ &
$-5.08496946 i$ &
$-4.15918692 i$ &
$-4.19874786 i$  &
$-4.21838805 i$  &
$\pm 0.93019026 - 3.53408745 i$ &
$\pm 0.39915578 - 1.78207262 i$    \\\hline

$\tilde{\omega} (l=1)$ &
$-6.13474436 i$ &
$-5.27886988 i$ &
$-5.36433739 i$  &
$\pm 1.13562848 - 4.81513725 i$  &
$-4.28198250 i$  &
$\pm 0.38802106 - 2.25895611 i$ \\\hline\hline

$\tilde{\omega} (l=2)$ &
$-1.99896780 i$ &
$\pm 4.26981231 - 1.26028729 i$  &
$\pm 3.26371359 - 0.97444667 i$ &
$\pm 2.84976486 - 0.85557590 i$  &
$\pm 2.12885667 - 0.64506380 i$  &
$\pm 0.77531941 - 0.23506578 i$   \\\hline

$\tilde{\omega} (l=2)$ &
$\pm 9.65211077 - 2.78317616 i$ &
$-1.99595932 i$ &
$-1.99424597 i$  &
$-1.99329952 i$  &
$\pm 2.02762970 - 1.96098183 i$  &
$\pm 0.76273725 - 0.70646734 i$    \\\hline

$\tilde{\omega} (l=2)$ &
$-4.00897172 i$ &
$\pm 3.93297577 - 3.86202953 i$ &
$\pm 3.04769906 - 2.97226270 i$  &
$\pm 2.67968181 - 2.60570195 i$  &
$-1.99126148 i$  &
$\pm 0.73662137 - 1.18285030 i$  \\\hline

$\tilde{\omega} (l=2)$ &
$-5.02743297 i$ &
$-4.03347941 i$ &
$-4.04665462 i$  &
$-4.05405720 i$  &
$\pm 1.85292830 - 3.32763153 i$  &
$\pm 0.69762034 - 1.67309576 i$    \\\hline

$\tilde{\omega} (l=2)$ &
$-6.05413248 i$ &
$-5.09907861 i$ &
$\pm 2.64644112 - 5.12013684 i$  &
$\pm 2.36287519 - 4.46119856 i$  &
$-4.06889645 i$  &
$-1.98621495 i$  \\\hline

$\tilde{\omega} (l=2)$ &
$-7.08840974 i$ &
$-6.18973980 i$ &
$-5.13697095 i$  &
$-5.15597703 i$  &
$\pm 1.60520496 - 4.77854571 i$  &
$\pm 0.66561953 - 2.18641889 i$  \\\hline

\end {tabular}
}
\end{table}\leavevmode\newline

\begin{table}[H]
\caption {Quasinormal frequencies $\tilde{\omega}$ for massive scalar fields in the background of Weyl black holes, with $\tilde{Q}= 0.5$. Here, the QNFs are obtained via the pseudospectral Chebyshev method using a number of Chebyshev polynomials in the range 95-100, 
with eight decimals places of accuracy for the QNF.}
\label{DominanceMassive1}\centering
\scalebox{0.65} {

\begin {tabular} { | c | c | c | c | c | c | c |}
\hline
${}$ & $\tilde{m} = 0$ & $ \tilde{m} = 1.0$  & $ \tilde{m} = 1.5$  &  $ \tilde{m} = 2.0$  & $ \tilde{m} = 2.5$ & $ \tilde{m} = 3.0$ \\\hline
$\tilde{\omega} (l=0)$ &
$0$ &
$-0.35401920 i$ &
$\pm 0.95381894 - 1.27335978 i$  &
$\pm 1.36020560 - 0.92104378 i$  &
$\pm 1.75401578 - 0.82539885 i$  &
$\pm 2.12144708 - 0.77631198 i$ 
 \\\hline
$\tilde{\omega} (l=0)$ &
$\pm 0.96483271 - 1.55503332 i$ &
$\pm 1.01952707 - 1.47518178 i$ &
$\pm 0.13490752 - 1.74387624 i$  &
$\pm 1.20505328 - 2.05701405 i$  &
$\pm 1.66267856 - 2.11130330 i$  &
$\pm 2.06533303 - 2.12576987 i$  \\\hline
$\tilde{\omega} (l=0)$ &
$-2.20642433 i$ &
$-2.53697632 i$ &
$-3.63647842 i$  &
$\pm 1.13901985 - 3.63088662 i$  &
$\pm 1.62955489 - 3.59059767 i$  &
$\pm 2.04711583 - 3.57066480 i$   \\\hline
$\tilde{\omega} (l=0)$ &
$-3.28026122 i$ &
$-3.02091679 i$ &
$-3.92292329 i$  &
$\pm 1.24105847 - 4.92822203 i$  &
$\pm 1.66347956 - 4.94973465 i$  &
$\pm 2.05830232 - 4.95625681 i$    \\\hline
$\tilde{\omega} (l=0)$ &
$-4.43842018 i$ &
$\pm 0.64770514 - 4.56196971 i$ &
$\pm 0.73302630 - 4.82169214 i$  &
$\pm 1.07793582 - 6.36649261 i$  &
$\pm 1.59110405 - 6.37295729 i$  &
$\pm 2.02191104 - 6.37435406 i$    \\\hline
$\tilde{\omega} (l=0)$ &
$\pm 0.65424586 - 4.76952419 i$ &
$-5.30401863 i$ &
$-6.09177144 i$  &
$\pm 1.20135686 - 7.83193696 i$  &
$\pm 1.63877715 - 7.80517185 i$  &
$\pm 2.04152319 - 7.79494730 i$  \\\hline\hline

$\tilde{\omega} (l=1)$ &
$-0.99061127 i$ &
$\pm 2.57340228 - 1.28359171 i$ &
$\pm 2.64781892 - 1.23549119 i$  &
$\pm 2.75809639 - 1.16893150 i$  &
$\pm 2.91051881 - 1.09024907 i$  &
$\pm 3.10582853 - 1.01171457 i$     \\\hline

$\tilde{\omega} (l=1)$ &
$\pm 2.51581819 - 1.32172179 i$ &
$-1.37985286 i$ &
$\pm 0.02364769 - 2.53655476 i$  &
$\pm 1.36864413 - 2.49182390 i$  &
$\pm 2.00214860 - 2.43802104 i$  &
$\pm 2.49496583 - 2.38927098 i$    \\\hline

$\tilde{\omega} (l=1)$ &
$-3.05278804 i$ &
$-3.49374404 i$ &
$\pm 2.07752476 - 4.01037748 i$  &
$\pm 2.12442253 - 3.89666887 i$  &
$\pm 2.32719802 - 3.78654458 i$  &
$\pm 2.61889124 - 3.73147422 i$    \\\hline

$\tilde{\omega} (l=1)$ &
$\pm 2.08871359 - 4.10594740 i$ &
$-3.72467809 i$ &
$-4.67287107 i$  &
$\pm 1.41898381 - 4.84547011 i$  &
$\pm 2.00029396 - 4.95852141 i$  &
$\pm 2.43806019 - 5.01559170 i$     \\\hline

$\tilde{\omega} (l=1)$ &
$-4.15918692 i$ &
$\pm 2.08283694 - 4.06686536 i$ &
$-4.81305355 i$  &
$\pm 1.51920174 - 6.67834041 i$  &
$\pm 2.00766689 - 6.54771666 i$  &
$\pm 2.41726580 - 6.49861598 i$     \\\hline

$\tilde{\omega} (l=1)$ &
$-5.27886988 i$ &
$-5.68024870 i$ &
$\pm 0.13477258 - 6.99707989 i$  &
$\pm 1.64821216 - 7.65109978 i$  &
$\pm 2.02975018 - 7.77124247 i$  &
$\pm 2.40669618 - 7.81516066 i$ \\\hline\hline

$\tilde{\omega} (l=2)$ &
$\pm 4.26981231 - 1.26028729 i$ &
$\pm 4.31054575 - 1.24622395 i$ &
$\pm 4.36160378 - 1.22887773 i$  &
$\pm 4.43334902 - 1.20505833 i$  &
$\pm 4.52603518 - 1.17530696 i$  &
$\pm 4.63993845 - 1.14043765 i$  \\\hline

$\tilde{\omega} (l=2)$ &
$-1.99595932 i$ &
$-2.38048725 i$ &
$-3.51812090 i$  &
$\pm 1.37043855 - 3.50803405 i$  &
$\pm 2.07546532 - 3.48850875 i$  &
$\pm 2.70741305 - 3.47046172 i$   \\\hline

$\tilde{\omega} (l=2)$ &
$\pm 3.93297577 - 3.86202953 i$ &
$\pm 3.95215272 - 3.82891733 i$ &
$-3.53147971 i$  &
$\pm 4.00979333 - 3.72741668 i$  &
$\pm 4.05226937 - 3.64939667 i$  &
$\pm 4.09922005 - 3.55177204 i$  \\\hline

$\tilde{\omega} (l=2)$ &
$-4.03347941 i$ &
$-4.44262643 i$ &
$\pm 3.97617699 - 3.78701034 i$  &
$\pm 1.44516766 - 5.63669615 i$  &
$\pm 2.18638626 - 5.60163478 i$  &
$\pm 2.81603316 - 5.52360166 i$   \\\hline

$\tilde{\omega} (l=2)$ &
$-5.09907861 i$ &
$-4.69417428 i$ &
$\pm 0.01894009 - 5.65239540 i$  &
$\pm 3.31833103 - 6.63366777 i$  &
$\pm 3.32368952 - 6.58696878 i$  &
$\pm 3.38491495 - 6.56813251 i$ \\\hline

$\tilde{\omega} (l=2)$ &
$-6.18973980 i$ &
$-6.61649306 i$ &
$\pm 3.32400665 - 6.67721158 i$  &
$\pm 1.46843955 - 7.86589037 i$  &
$\pm 2.18281358 - 7.87734320 i$  &
$\pm 2.75163403 - 7.90674509 i$ \\\hline

\end {tabular}
}
\end{table}\leavevmode\newline

\begin{table}[H]
\caption {Quasinormal frequencies $\tilde{\omega}$ for massive scalar fields in the background of Weyl black holes, with $\tilde{Q}= 0.75$. Here, the QNFs are obtained via the pseudospectral Chebyshev method using a number of Chebyshev polynomials in the range 95-100, 
with eight decimals places of accuracy for the QNF.}
\label {DominanceMassive2}\centering
\scalebox{0.65} {

\begin {tabular} { | c | c | c | c | c | c | c |}
\hline
${}$ & $ \tilde{m} = 0$ & $ \tilde{m} = 1.0$  & $ \tilde{m} = 1.5$  &  $ \tilde{m} = 2.0$  & $ \tilde{m} = 2.5$ & $ \tilde{m} = 3.0$ \\\hline
$\tilde{\omega} (l=0)$ &
$0$ &
$-0.29986955 i$ &
$\pm 0.60482925 - 0.61017470 i$  &
$\pm 0.91330080 - 0.54193701 i$  &
$\pm 1.18511733 - 0.51897328 i$  &
$\pm 1.44617318 - 0.50838546 i$ 
 \\\hline
$\tilde{\omega} (l=0)$ &
$\pm 0.46343497 - 0.93126623 i$ &
$\pm 0.47960014 - 0.88467863 i$ &
$-1.43086052 i$  &
$\pm 0.81984856 - 1.53466704 i$  &
$\pm 1.14858827 - 1.51850535 i$  &
$\pm 1.43116695 - 1.50879251 i$  \\\hline
$\tilde{\omega} (l=0)$ &
$-2.21473623 i$ &
$\pm 0.43234453 - 2.13615514 i$ &
$-1.76138964 i$  &
$\pm 0.83584307 - 2.49291707 i$  &
$\pm 1.14870871 - 2.50297651 i$  &
$\pm 1.43043159 - 2.50260783 i$   \\\hline
$\tilde{\omega} (l=0)$ &
$\pm 0.33711158 - 2.31236354 i$ &
$-2.73424385 i$ &
$\pm 0.39430601 - 2.39664359 i$  &
$\pm 0.81839997 - 3.52753666 i$  &
$\pm 1.14175680 - 3.51210914 i$  &
$\pm 1.42778669 - 3.50574235 i$    \\\hline
$\tilde{\omega} (l=0)$ &
$-3.41123404 i$ &
$-3.17994322 i$ &
$\pm 0.28627815 - 3.64178423 i$  &
$\pm 0.79158397 - 4.49286445 i$  &
$\pm 1.13108718 - 4.50186364 i$  &
$\pm 1.42315664 - 4.50243394 i$    \\\hline
$\tilde{\omega} (l=0)$ &
$\pm 0.32713306 - 3.94683557 i$ &
$\pm 0.42220511 - 3.96456323 i$ &
$-4.22244761 i$  &
$\pm 0.81978123 - 5.50449805 i$  &
$\pm 1.13668525 - 5.50386936 i$  &
$\pm 1.42413904 - 5.50275080 i$  \\\hline\hline

$\tilde{\omega} (l=1)$ &
$\pm 1.195576720 - 0.693686395 i$ &
$\pm 1.27834073 - 0.65597656 i$ &
$\pm 1.38369056 - 0.62081840 i$  &
$\pm 1.52560847 - 0.58803248 i$  &
$\pm 1.69438861 - 0.56210273 i$  &
$\pm 1.88199643 - 0.54311579 i$     \\\hline

$\tilde{\omega} (l=1)$ &
$-0.98353510 i$ &
$-1.37455031 i$ &
$\pm 1.17283467 - 1.90769927 i$  &
$\pm 1.34594677 - 1.72597668 i$  &
$\pm 1.57822699 - 1.64013200 i$  &
$\pm 1.80810711 - 1.59400973 i$    \\\hline

$\tilde{\omega} (l=1)$ &
$\pm 1.13539826 - 2.08983835 i$ &
$\pm 1.15531011 - 2.03138306 i$ &
$-2.50857981 i$  &
$\pm 1.11585411 - 2.58836748 i$  &
$\pm 1.46199981 - 2.58534673 i$  &
$\pm 1.73657392 - 2.57273034 i$    \\\hline

$\tilde{\omega} (l=1)$ &
$-3.10011501 i$ &
$\pm 0.87558647 - 3.46770947 i$ &
$-2.64119299 i$  &
$\pm 1.17578965 - 3.54755095 i$  &
$\pm 1.44913713 - 3.55758125 i$  &
$\pm 1.71076159 - 3.55539213 i$     \\\hline

$\tilde{\omega} (l=1)$ &
$\pm 0.93019026 - 3.53408745 i$ &
$-3.59806840 i$ &
$\pm 0.91695132 - 3.48988025 i$  &
$\pm 1.06014894 - 4.57995989 i$  &
$\pm 1.39420762 - 4.55750026 i$  &
$\pm 1.67598645 - 4.54934445 i$     \\\hline

$\tilde{\omega} (l=1)$ &
$-4.28198250 i$ &
$-3.87029770 i$ &
$\pm 0.13795541 - 4.86092104 i$  &
$\pm 1.05297011 - 5.49634089 i$  &
$\pm 1.37274946 - 5.52676056 i$  &
$\pm 1.65448807 - 5.53397918 i$ \\\hline\hline

$\tilde{\omega} (l=2)$ &
$\pm 2.12885667 - 0.64506380 i$ &
$\pm 2.17925659 - 0.63453647 i$ &
$\pm 2.24168764 - 0.62260830 i$  &
$\pm 2.32779988 - 0.60800547 i$  &
$\pm 2.43597968 - 0.59232874 i$  &
$\pm 2.56405464 - 0.57698352 i$  \\\hline

$\tilde{\omega} (l=2)$ &
$\pm 2.02762970 - 1.96098183 i$ &
$\pm 2.06804196 - 1.92630590 i$ &
$\pm 2.12272325 - 1.88345614 i$  &
$\pm 2.20682283 - 1.82781718 i$  &
$\pm 2.32276837 - 1.76854169 i$  &
$\pm 2.46614272 - 1.71465469 i$   \\\hline

$\tilde{\omega} (l=2)$ &
$-1.99126148 i$ &
$-2.37876245 i$ &
$\pm 1.87473675 - 3.21778921 i$  &
$\pm 1.88282926 - 3.07163189 i$  &
$\pm 2.04783511 - 2.87110193 i$  &
$\pm 2.27072005 - 2.76636604 i$  \\\hline

$\tilde{\omega} (l=2)$ &
$\pm 1.85292830 - 3.32763153 i$ &
$\pm 1.86577948 - 3.28176135 i$ &
$-3.53661357 i$  &
$\pm 1.37709735 - 3.57039692 i$  &
$\pm 1.84512589 - 3.65259917 i$  &
$\pm 2.14005594 - 3.66595160 i$   \\\hline

$\tilde{\omega} (l=2)$ &
$-4.06889645 i$ &
$-4.50772208 i$ &
$-3.56589860 i$  &
$\pm 1.67660888 - 4.66992099 i$  &
$\pm 1.87552083 - 4.65207989 i$  &
$\pm 2.10829622 - 4.64182895 i$ \\\hline

$\tilde{\omega} (l=2)$ &
$\pm 1.60520496 - 4.77854571 i$ &
$\pm 1.58819088 - 4.74940080 i$ &
$\pm 1.58454997 - 4.71039303 i$  &
$\pm 1.34339703 - 5.66426999 i$  &
$\pm 1.73679933 - 5.62798195 i$  &
$\pm 2.02881589 - 5.61791155 i$ \\\hline

\end {tabular}
}
\end{table}\leavevmode\newline

\end{document}